\documentclass[lettersize,journal]{IEEEtran}

\usepackage{amsmath,amsfonts}
\usepackage{algorithmic}
\usepackage[ruled]{algorithm2e}
\usepackage{array}
\usepackage{amssymb} % triangleq
\usepackage[caption=false,font=normalsize,labelfont=sf,textfont=sf]{subfig}
\usepackage{textcomp}
\usepackage{stfloats}
\usepackage{url}
\usepackage{verbatim} % multiline comment
\usepackage{graphicx}
\usepackage{epstopdf}
\usepackage{cite}
\def\BibTeX{{\rm B\kern-.05em{\sc i\kern-.025em b}\kern-.08em
    T\kern-.1667em\lower.7ex\hbox{E}\kern-.125emX}}
\usepackage{balance}
\usepackage{cleveref}

\IEEEoverridecommandlockouts

\begin{document}

\title{Movable Antenna Aided NOMA: Joint Antenna Positioning, Precoding, and Decoding Design}
\author{Zhenyu Xiao, ~\IEEEmembership{Senior Member,~IEEE,}
	Zhe Li, ~\IEEEmembership{Graduate Student Member,~IEEE,}
	Lipeng Zhu, ~\IEEEmembership{Member,~IEEE,} 
	Boyu Ning,~\IEEEmembership{Member,~IEEE,}
	Daniel Benevides da Costa, ~\IEEEmembership{Senior Member,~IEEE,} Xiang-Gen Xia,~\IEEEmembership{Fellow,~IEEE,}
	and Rui Zhang,~\IEEEmembership{Fellow,~IEEE}
\vspace{-0.8 cm}
\thanks{This work was supported in part by the National Natural Science Foundation of China (NSFC) under grant numbers 62171010 and U22A2007, the Beijing Natural Science Foundation under grant number L212003, and the Fundamental Research Funds for the Central Universities under grant number YWF-22-JC-10. (\emph{Corresponding author: Lipeng Zhu})}
\thanks{Z. Xiao and Z. Li are with the School of Electronic and Information Engineering, Beihang University, Beijing 100191, China. (e-mail: xiaozy@buaa.edu.cn, zheli@buaa.edu.cn).}

\thanks{L. Zhu is with the Department of Electrical and Computer Engineering, National University of Singapore, Singapore 117583, Singapore. (e-mail: zhulp@nus.edu.sg).}

\thanks{B. Ning is with National Key Laboratory of Science and Technology on Communications, University of Electronic Science and Technology of China (UESTC), Chengdu 611731, China (e-mail: boydning@outlook.com).}

\thanks{D. B. da Costa is with the Department of Electrical Engineering, King Fahd University of Petroleum \& Minerals (KFUPM), Dhahran 31261, Saudi Arabia (email: danielbcosta@ieee.org).}

\thanks{X.-G. Xia is with the Department of Electrical and Computer Engineering, University of Delaware, Newark, DE 19716, USA. (e-mail: xxia@ee.udel.edu).}

\thanks{R. Zhang is with School of Science and Engineering, Shenzhen Research Institute of Big Data, The Chinese University of Hong Kong, Shenzhen, Guangdong 518172, China (e-mail:rzhang@cuhk.edu.cn). He is also with the Department of Electrical and Computer Engineering, National University of Singapore, Singapore 117583 (e-mail: elezhang@nus.edu.sg).}}

%\date{February 2024}

% \markboth{Journal of \LaTeX\ Class Files,~Vol.~18, No.~9, September~2020}%
% {How to Use the IEEEtran \LaTeX \ Templates}

\maketitle

\begin{abstract}
This paper investigates movable antenna (MA) aided non-orthogonal multiple access (NOMA) for multi-user downlink communication, where the base station (BS) is equipped with a fixed-position antenna (FPA) array to serve multiple MA-enabled users. An optimization problem is formulated to maximize the minimum achievable rate among all the users by jointly optimizing the MA positioning of each user, the precoding matrix at the BS, and the successive interference cancellation (SIC) decoding indicator matrix at the users, subject to a set of constraints including the limited movement area of the MAs, the maximum transmit power of the BS, and the SIC decoding condition. To solve this non-convex problem, we propose a two-loop iterative optimization algorithm that combines the hippopotamus optimization (HO) method with the alternating optimization (AO) method to obtain a suboptimal solution efficiently. Specifically, in the inner loop, the complex-valued precoding matrix and the binary decoding indicator matrix are optimized alternatively by the successive convex approximation (SCA) technique with customized greedy search to maximize the minimum achievable rate for the given positions of the MAs. In the outer loop, each user's antenna position is updated using the HO algorithm, following a novel nature-inspired intelligent optimization framework. Simulation results show that the proposed algorithms can effectively avoid local optimum for highly coupled variables and significantly improve the rate performance of the NOMA system compared to the conventional FPA system as well as other benchmark schemes.
\end{abstract}
\vspace{-0.1 cm}
\vspace{-0.5 cm}
\section{Introduction}
The sixth-generation (6G) wireless communication networks are expected to support high data rate, enhanced reliability, reduced latency, and energy-efficient operations \cite{1wang2023road}, \cite{ 2saad2019vision}. Owing to resource constraints in time/frequency/space domain, traditional orthogonal multiple access (OMA) techniques, such as frequency division multiple access (FDMA), time division multiple access (TDMA), code division multiple access (CDMA), and orthogonal frequency division multiple access (OFDMA), are challenging to fulfill the requirements of the increasing throughput in contemporary networks \cite{3zhu2019joint}, \cite{4xiao2018joint}. Therefore, it is essential to design efficient multiple-access techniques to manage the surge in data traffic for the next-generation wireless communication systems \cite{5dai2015non}. Non-orthogonal multiple access (NOMA) technique is capable of achieving high spectral and energy efficiency, which can cater for dense users in resource-limited scenarios \cite{5dai2015non, 6li2021joint, 7zhao2019secure}. Specifically, multiple users share the same time-frequency resource block yet with different power levels allocated to distinguish between them. In particular for downlink NOMA, the transmitted symbols for multiple users are superimposed and coded at the baseband. At the receiver's end, the signals are decoded using the successive interference cancellation (SIC) technique \cite{8benjebbour2013concept}. As such, this technology can notably enhance spectral efficiency in multi-user communications and alleviate the issue of the increasingly scarce spectrum resources \cite{9pervez2021joint}. To further improve the communication performance to meet the higher requirements of 6G networks, the spatial resources should be fully exploited. To this end, multiple-input multiple-output (MIMO) systems and array beamforming technologies are essential for achieving spatial multiplexing gains in future wireless communication \cite{10ning2023beamforming}, \cite{11paulraj2004overview}, and the MIMO-NOMA has been shown to have a significant improvement in the achievable rate for multi-user uplink and downlink communications \cite{12krishnamoorthy2021uplink}.

However, conventional MIMO systems usually adopt fixed-position antenna (FPA) arrays, which cannot change their positions to fully exploit the spatial degrees of freedom (DoFs) and spatial diversity/multiplexing gains \cite{13zhu2023movable}, \cite{14xiao2019user}. To address this issue, an innovative antenna architecture called movable antenna (MA) has recently been proposed \cite{15zhu2023modeling}, which is also known as fluid antenna system (FAS) in terms of flexible antenna positioning \cite{16zhu2024historical}. With the aid of driving components such as a motor, the MA can flexibly adjust its position and/or orientation in three-dimensional (3D) space instead of being constrained to some preset positions/orientations, thus fully exploiting the potential of spatial diversity, beamforming, and multiplexing gains \cite{15zhu2023modeling,zhu2024wideband,17zhuMABeamforming,zhu2024dynamic,27ma2023mimo,18caoMAcompress}. Specific implementation architectures of MAs can be referred to \cite{13zhu2023movable,ning2024movable}.

Previous studies have demonstrated the advantages of MA-aided communication over conventional FPA communication systems. 
 In \cite{13zhu2023movable}, the concept and challenges of MA in wireless communications were delineated, where the hardware systems and channel models of MA were then demonstrated. 
 In \cite{15zhu2023modeling}, a multi-path MA channel model based on the field response was developed. Then, the performance of the MA system was analyzed under deterministic and stochastic channel conditions. 
 In \cite{19mei2024movable}, an equivalent fixed-hop shortest path approach in graph theory was applied to solve discrete position optimization problem for MAs, allowing MA-enabled multiple-input single-output (MISO) communication systems to achieve higher received signal power. 
 The authors in \cite{20wei2024joint} applied MAs to cognitive radio (CR) networks. By optimizing the positions of MAs, the spatial diversity gains were fully exploited, enhancing the received signal power at the secondary receiver while suppressing interference from primary receivers.
 The studies in \cite{21hu2024movable} explored the use of MAs in wireless communication systems with coordinated multi-point (CoMP) reception. For this setup, the effective received signal-to-noise ratio (SNR) between the transmitter and multiple destinations was maximized. The results demonstrated that MAs can achieve higher SNR, thereby enhancing the quality and reliability of the communication system. 
 In \cite{22gao2024joint}, MAs were used to enhance the weighted minimum signal-to-noise-plus-interference ratio (SINR) among users in a multicast communication scenario. 
 In \cite{23wang2024movable}, the total transmit power of the interference network was suppressed through the extra spatial DoFs provided by the MAs, enabling the desired signal to achieve more reliable reception. 
%The studies in \cite{19mei2024movable,20wei2024joint,21hu2024movable,22gao2024joint,23wang2024movable} applied MAs to achieve higher received signal-to-noise ratio (SNR) by fully exploiting spatial diversity gains, thereby enhancing the quality and reliability of the communication systems. 
The authors in \cite{24hu2024movable} and \cite{25tang2024secure} investigated the secrecy outage probability and the secrecy rate in secure communication scenarios, respectively. The results showed that MA-assisted communication systems can achieve better secrecy performance when eavesdroppers were present. 
In \cite{26wu2024movable}, MAs were introduced to assist intelligent reflecting surface (IRS)/reconfigurable intelligent surface (RIS)-aided sensing and  communication systems by further exploiting the spatial diversity gain. The results indicated that targets located in non-line-of-sight (NLoS) areas can be well detected and the beampattern gain towards them can be enhanced, which can improve communication performance in complex multi-path scenarios.

For the enhancement of spatial multiplexing gain by using MAs, the channel model of the MIMO communication system with MAs in \cite{27ma2023mimo} was mathematically described, and its performance was characterized. 
The uplink scenario of the MA-aided multi-user communication was discussed in \cite{28zhu2023movable}, where the multiple access channel was considered. Furthermore, the receive combining matrix was optimized using the minimum mean square error (MMSE) and zero-forcing (ZF) combining methods. Then, the positions of MAs were optimized efficiently by using a combination of the gradient descent method, the gradient descent with momentum method, and the descent direction of the quasi-Newton’s method to jointly obtain an optimal descent direction. The results showed that a significant power-saving uplink multi-user communication can be achieved by MA position optimization.
%The uplink scenario of the MA-aided multi-user communication was discussed in \cite{28zhu2023movable}, where the multiple access channel was considered. Then, the positions of MAs were optimized efficiently by using a combination of the gradient descent method, the gradient descent with momentum method, and the descent direction of the quasi-Newton’s method to jointly obtain an optimal descent direction. The results showed that a significant power-saving uplink multi-user communication can be achieved by MA position optimization.
In \cite{29qin2023antenna}, the MA-aided multi-user downlink MISO communication was investigated. An alternating optimization (AO) algorithm based on successive convex approximation (SCA) and penalty method was proposed. Numerical results showed that in the downlink scenario, the MA-aided communication system can yield a higher energy efficiency than the FPA communication system. 
In \cite{30pi2023multiuser}, the MAs were deployed at base station (BS). To improve the minimum achievable rate among users, a two-loop iterative approach based on the particle swarm optimization (PSO) method as well as the block coordinate descent (BCD) method was proposed to optimize MAs’ positions, receive combining matrix and power allocation.

The aforementioned MA-related works on multiuser communications \cite{28zhu2023movable,29qin2023antenna,30pi2023multiuser} mainly focus on space division multiple access (SDMA) techniques, while SDMA cannot fully leverage the advantages of MA due to limited spectrum efficiency. In contrast, NOMA can make full use of limited communication resources in multi-user communication scenarios by differentiating power allocation among users. Therefore, the performance analysis of NOMA communication systems assisted by MA is an area that still requires investigation. In \cite{31zhou2024movable}, a two-user MA-assisted NOMA downlink communication system was investigated. The results showed that NOMA can achieve higher achievable rate compared to OMA systems by jointly optimizing power allocation and MA positioning. An MA-aided short-packet communication system with two-user NOMA was introduced in \cite{32he2024movable}. By optimizing antenna positions, the effective throughput of the core user was improved. In \cite{33li2024sum}, an MA-enabled uplink NOMA communication system was considered to maximize the sum rate by the joint optimization of MAs' positions, decoding order, and power allocation. The results indicated that the MA-NOMA system can deliver better rate performance compared to both FPA systems and OMA systems.
Inspired by the above, in this paper, we study a new setup of  MA-aided multi-user NOMA systems, where each user is equipped with a single-MA while the BS is equipped with the conventional fixed-position uniform planar array (UPA). Users here are, for example, Internet of Things (IoT) devices or machine-type communication (MTC) devices, which are characterized by low mobility and have sufficient internal space to deploy MAs, thereby reducing the burden on deploying MAs at the BS. To fully unleash the potential of MA-NOMA systems, we design an adaptive precoding and decoding scheme to adapt to the correlation between channel vectors for multiple users, which distinguishes with the existing works on MA-NOMA without precoding or with fixed decoding strategies \cite{31zhou2024movable,32he2024movable,33li2024sum}. The main contributions of this paper are summarized as follows:

\begin{enumerate}
    % 添加第一个项目
    \item We propose to deploy a single-MA at each user to improve the NOMA-based multi-user downlink communication system. The broadcast channel (BC) is modeled as a function with respect to the antenna position vector (APV) to characterize the multi-path response. Based on this channel model, we formulate an optimization problem to maximize the minimum achievable rate by jointly optimizing the MAs’ positions, the precoding matrix as well as the decoding indicator matrix, subject to the constraints of finite movable area of the MAs, maximum transmit power of the BS, and the SIC decoding condition.
    % 添加第二个项目
    \item To solve the formulated non-convex problem, a two-loop iterative optimization algorithm incorporating the hippopotamus optimization (HO) method and the AO method is employed to obtain a suboptimal solution efficiently. In the inner loop, for any given APV, the precoding matrix and decoding indicator matrix are optimized iteratively by the AO method to maximize the minimum achievable rate. In particular, the SCA technique is used to transform the non-convex problem into a convex one and optimize the precoding matrix of the BS. Moreover, a customized greedy search method is implemented to obtain a suboptimal decoding indicator matrix. In the outer loop, the APVs are optimized by a nature-inspired HO algorithm, where each hippopotamus represents an APV and the fitness is defined as the minimum achievable rate among users.

    \item Extensive simulation results show that optimizing the positions of the MAs enhances the rate performance of the MA-aided communication systems compared to the conventional systems utilizing FPAs. The simulation results also confirm that our proposed NOMA scheme with adaptive decoding has better rate performance compared to SDMA scheme and conventional NOMA scheme with fixed decoding order for varying number of antennas and users, due to the fact that our proposed solution can more efficiently mitigate the inter-user interference and achieve greater adaptability. In addition, we evaluate the performance of the proposed algorithm under imperfect field-response information (FRI), which shows that our proposed algorithm remains robust in the presence of misestimated angles and response coefficients for the channel paths.

\end{enumerate}

The rest of this paper is organized as follows. In Section II, we introduce the system model for MA-aided NOMA communication systems and formulate the optimization problem to maximize the minimum achievable rate. In Section III, we present the solution for the formulated optimization problem. In Section IV, we show the simulation results. Finally, conclusions are provided in Section V.

\textit{Notation}: $a$, $\bf a$, $\bf A$ and $\mathcal{A}$ represent a scalar, a vector, a matrix and a set, respectively. $\left( \cdot \right)^{\text{T}}$ and $\left( \cdot \right)^{\text{H}}$ denote transpose and conjugate transpose, respectively. $\left| \cdot \right|$ and $\left\| \cdot \right\|_2$ denote the absolute value and Euclidean norm, respectively. $\left[ {\bf a} \right]_i$ and $\left[ {\bf A} \right]_{i,j}$ denote the $i$-th entry of vector $\bf a$ and the entry in the $i$-th row and $j$-th column of matrix $\bf A$, respectively. ${\rm tr}\left( {\bf A} \right)$ and ${\rm Re}\left( {\bf A} \right)$ denote the trace and the real part of matrix $\bf A$, respectively. ${\bf I}_N$ is the identical matrix of size $N \times N$  and ${\bf 1}_{N}$ denotes the $N$-dimensional column vector with all entries equal to one.  $\odot$ is the Hadamard product. $\mathcal{CN}{\left( {0}, {\sigma ^2}\right)}$ denotes the circularly symmetric complex Gaussian (CSCG) distribution with mean zero and variance $\sigma ^2$. $\mathcal{U}\left[ {a, b} \right]$ represents the uniform distribution between the real-number $a$ and $b$. $\mathbb{R}$ and $\mathbb{C}$ denote the sets of real and complex numbers, respectively. $\mathbb{E\{ \cdot \}}$ denotes the expected value of a random variable.

\section{System Model and Problem Formulation}
\begin{figure}[!t]
\centering
\includegraphics[width=3.5in]{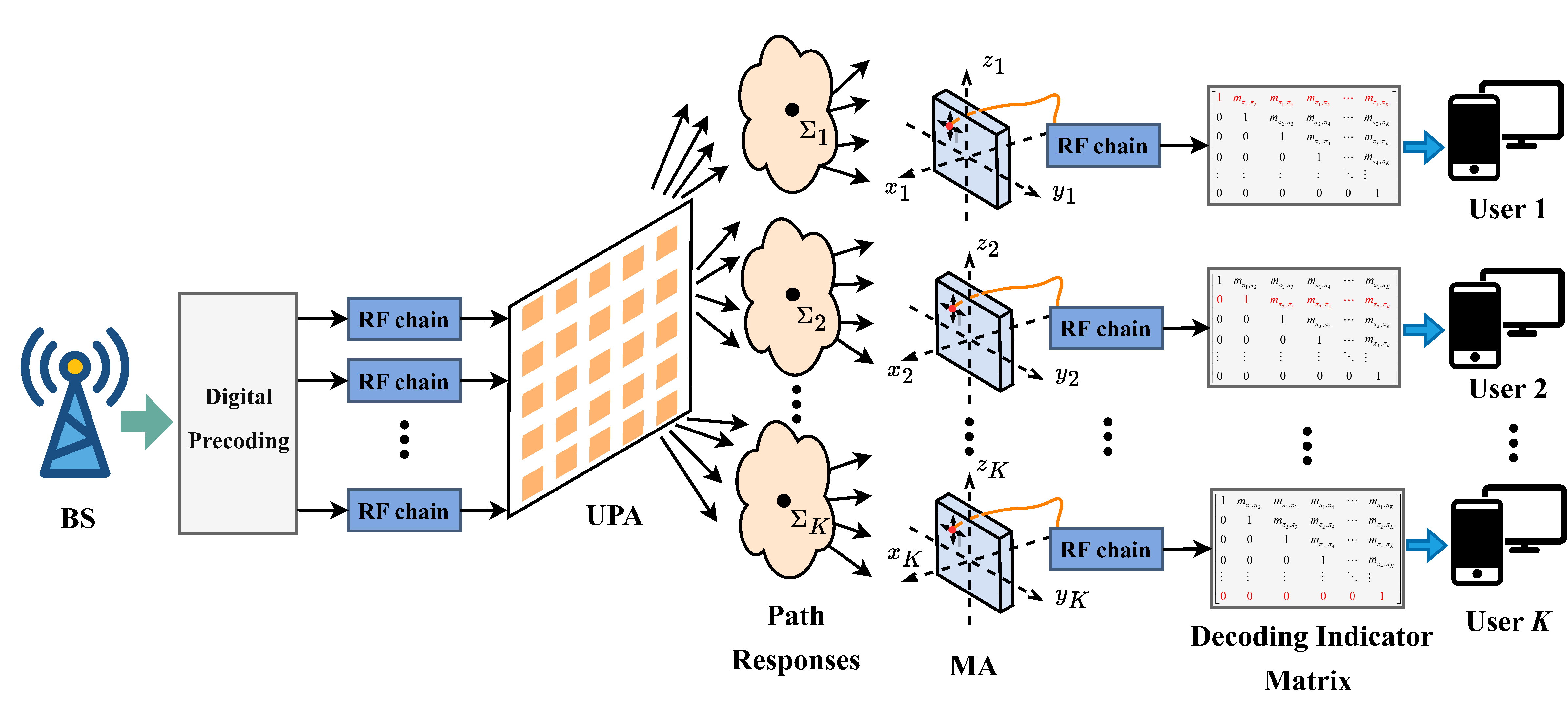}
\caption{Illustration of the downlink NOMA transmission with adaptive decoding between the BS and $K$ single-MA users.}
\label{fig: MA-aided system}
\end{figure}

As shown in Fig.~\ref{fig: MA-aided system}, we consider a downlink multi-user communication system with $K$ users, indexed by $\mathcal{K}=\{1,2,...,K\}$. The $K$ single-MA users are served by the BS equipped with a fixed-position UPA of size $N = N_1 \times N_2$, with $N_1$ and $N_2$ denoting the numbers of antennas along horizontal and vertical directions, respectively. For each user $k$, the single-MA is connected to the radio frequency chain via a flexible cable so that it can move freely within a 3D local region $\mathcal{C}_k$. Without loss of generality, the local position coordinate of the MA for user $k$ with respect to the origin $O_k$  of the movable region is denoted as ${{\bf u}_k} = {\left[{x_k},{y_k},{z_k}\right]^{\rm T}} \in {{\cal C}{_k}}, 1 \le k \le K$. The 3D movable region of the MA is assumed to be a cube of size $\left[- \frac{A}{2},\frac{A}{2}\right] \times \left[ - \frac{A}{2},\frac{A}{2}\right] \times \left[ - \frac{A}{2},\frac{A}{2}\right]$. In addition, the local position coordinate of the $n$-th FPA at the BS is represented as ${{\bf v}_n} = {\left[{X_n},{Y_n},{Z_n}\right]^{\rm T}}, 1 \le n \le N$.

The channel vector from the BS to user $k$ is denoted by
${{\bf h}_k \left( {\bf u}_k \right)} \in {\mathbb{C}^{{N} \times 1}}$, which is determined by the propagation environment and the position of the MA \cite{15zhu2023modeling}, \cite{28zhu2023movable}. Then, the received signals at all the $K$ users can be expressed as          
\begin{equation}
\label{eq: receive_signal1}
{\bf{y}} = {{\bf{H}}^\mathrm{H}}\left({\bf{\tilde u}}\right){\bf{Ws}} + {\bf{n}},
\end{equation}
where ${{\bf{H}}}({\bf{\tilde u}})=[{{\bf{h}}_1}({{\bf{u}}_1}),{{\bf{h}}_2}({{\bf{u}}_2}),...,{{\bf{h}}_K}({{\bf{u}}_K})] \in {\mathbb{C}^{{N} \times K}}$ is the channel matrix from the UPA at the BS to all $K$ users, and ${\bf{\tilde u}} = {\left[{{\bf{u}}_1^\mathrm{T}},{{\bf{u}}_2^\mathrm{T}},...,{{\bf{u}}_K}^\mathrm{T}\right]^\mathrm{T}} \in {\mathbb{R}^{3K \times 1}}$ is the APV which contains the MAs' position coordinates for all users. ${\bf{W}} = \left[{{\bf{w}}_1},{{\bf{w}}_2},...,{{\bf{w}}_K}\right] \in {\mathbb{C}^{{N} \times K}}$ is the digital precoding matrix of the BS, and ${\bf{w}}_k$ is the precoding vector for user $k$, $1 \le k \le K$. ${\bf{s}} = {[{s_1},{s_2},...,{s_K}]^\mathrm{T}} \in {\mathbb{C}^{K \times 1}}$ denotes the transmitted signals of users with normalized power, i.e., $\mathbb{E}\{{\bf s}{\bf s}^\mathrm{H}\}={\bf{I}}_K$. ${\bf{n}} = [{n_1},{n_2},...,{n_K}]^\mathrm{T}\sim{{\cal C}{\cal N}}(0,{\sigma ^2}{{\bf{I}}_K})$ is the zero mean additive white Gaussian noise (AWGN) at users with average power $\sigma^ 2$. In particular, the received signal at user $k$ can be represented as
\begin{equation}
\label{eq: receive_signal2}
{y_k} = {\bf{h}}_k^{\rm{H}}({{\bf{u}}_k})\sum\limits_{i = 1}^K {{{\bf{w}}_i}{s_i} + {n_k}} ,1 \le k \le K.
\end{equation}

% \begin{figure*}[ht]
%     \centering
%     \includegraphics[width=\linewidth]{figures/MA-aided-system.png}
%     \caption{Illustration of the downlink NOMA transmission with adaptive decoding between the BS and $K$ single-MA users.}
%     \label{fig: MA-aided system}
% \end{figure*} 

% \begin{figure}[!t]
% \centering
% \includegraphics[width=3in]{figures/coordinate.eps}
% \caption{Illustration of the 3D local coordinate system and the corresponding spatial angles of arrival for user $k$, $1 \le k \le K$.}
% \label{fig: coordinate}
% \end{figure}

\subsection{Channel Model}
Similar to \cite{28zhu2023movable}, we consider narrow-band channels with flat fading and assume the far-field condition between the BS and users. In other words, the movable region $\mathcal C_k$ of each MA and the size of the UPA at the BS are much smaller than the propagation distance of the signal. Thus, we can consider that the angles of departure (AoDs), the angles of arrival (AoAs), and the amplitudes of the complex coefficients for channel paths remain constant within the movable area for each MA, while the phases of the channel paths change with the positions of the MAs.

Denote the total numbers of transmit and receive channel paths from the BS to user $k$ by $L^t_k$ and $L^r_k$, $1 \le k \le K$, respectively. The azimuth and elevation AoAs for the $j$-th receive path from the BS to user $k$ are denoted as $\phi^r_{k,j}$ and $\theta^r_{k,j}$, $1 \le j \le L^r_k$, respectively. The azimuth and elevation AoDs for the $i$-th transmit path from the BS to user $k$ are denoted as $\phi^t_{k,i}$ and $\theta^t_{k,i}$, $1 \le i \le L^t_k$, respectively. Further, to represent the spatial relations more conveniently, we introduce virtual AoAs and AoDs as, for $1 \le k \le K, 1 \le j \le L^r_k$, and $1 \le i \le L^t_k$,
\begin{subequations}%\label{eq: virtual angles}
\begin{align}
&\begin{cases}
\vartheta _{k,j}^r &= \cos \theta _{k,j}^r\cos \phi _{k,j}^r, \\
\varphi _{k,j}^r &= \cos \theta _{k,j}^r\sin \phi _{k,j}^r,\\
\omega _{k,j}^r &= \sin \theta _{k,j}^r,  \end{cases} \label{eq: virtual AOA} \\
&\begin{cases}
\vartheta _{k,i}^t &= \cos \theta _{k,i}^t\cos \phi _{k,i}^t, \\
\varphi _{k,i}^t &= \cos \theta _{k,i}^t\sin \phi _{k,i}^t,\\
\omega _{k,i}^t &= \sin \theta _{k,i}^t.  \end{cases} \label{eq: virtual AOD}
\end{align}
\end{subequations} 

%\label{eq: virtual AOA}
%\begin{subequations}%\label{eq: virtual angles}
%\begin{align}
%\begin{cases}
%\vartheta _{k,j}^r &= \cos \theta _{k,j}^r\cos \phi _{k,j}^r, {\rm{ }}\varphi _{k,j}^r = \cos \theta _{k,j}^r\sin \phi _{k,j}^r, \\
%{\rm{ }}\omega _{k,j}^r &= \sin \theta _{k,j}^r,\quad\quad\quad        1 \le k \le K,1 \le j \le L^r_k,  \end{cases} \label{eq: virtual AOA} \\
%\begin{cases}
%\vartheta _{k,i}^t &= \cos \theta _{k,i}^t\cos \phi _{k,i}^t, \:\: {\rm{ }}\varphi _{k,i}^t = \cos \theta _{k,i}^t\sin \phi _{k,i}^t, \\
%{\rm{ }}\omega _{k,i}^t &= \sin \theta _{k,i}^t,\quad\quad\quad\:        1 \le k \le K,1 \le i \le L^t_k.  \end{cases} \label{eq: virtual AOD}
%\end{align}
%\end{subequations} 

Then, the signal propagation distance difference of the $j$-th receive path for user $k$ between the position of the MA and the reference point (i.e., $O_k$) of the local coordinate is denoted as $\rho _{k,j}^r({{\bf{u}}_k})$, while the signal propagation distance difference of the $i$-th transmit path for user $k$ between the position of the $n$-th FPA of the UPA and the origin of the local coordinate at the BS is represented as $\rho_{k,i}^t({{\bf{v}}_n})$, which are given by
\begin{subequations}\label{eq: distance difference}
\begin{align}
\rho _{k,j}^r({{\bf{u}}_k}) &= {x_k}\vartheta _{k,j}^r + {y_k}\varphi _{k,j}^r + {z_k}\omega _{k,j}^r, \:\: 1 \le j \le L_k^r,\label{eq: receive difference}\\
\rho _{k,i}^t({{\bf{v}}_n}) &= {X_n}\vartheta _{k,i}^t + {Y_n}\varphi _{k,i}^t + {Z_n}\omega _{k,i}^t, 1 \le i \le L_k^t.\label{eq: transmit diffrence}
\end{align}
\end{subequations}

Then, the transmit and receive field-response vectors (FRVs) for the channel between the BS and user $k$ are obtained as \cite{15zhu2023modeling}
\begin{subequations}\label{eq: FRV}
\begin{align}
{{\bf{g}}_k}({{\bf{v}}_n}) &= {\left[{{\mathop{\rm e}\nolimits} ^{{\rm{j}}\frac{{2\pi }}{\lambda }\rho _{k,1}^t({{\bf{v}}_n})}},{{\mathop{\rm e}\nolimits} ^{{\rm{j}}\frac{{2\pi }}{\lambda }\rho _{k,2}^t({{\bf{v}}_n})}},...,{{\mathop{\rm e}\nolimits} ^{{\rm{j}}\frac{{2\pi }}{\lambda }\rho _{k,L_k^t}^t({{\bf{v}}_n})}}\right]^\mathrm{T}},\label{eq: transmit FRV}\\
{{\bf{f}}_k}({{\bf{u}}_k}) &= {\left[{{\mathop{\rm e}\nolimits} ^{{\rm{j}}\frac{{2\pi }}{\lambda }\rho _{k,1}^r({{\bf{u}}_k})}},{{\mathop{\rm e}\nolimits} ^{{\rm{j}}\frac{{2\pi }}{\lambda }\rho _{k,2}^r({{\bf{u}}_k})}},...,{{\mathop{\rm e}\nolimits} ^{{\rm{j}}\frac{{2\pi }}{\lambda }\rho _{k,L_k^r}^r({{\bf{u}}_k})}}\right]^\mathrm{T}} \: ,\label{eq: receive FRV}
\end{align}
\end{subequations}
with $1 \le k \le K$ and $1 \le n \le N$, where $\lambda$ is carrier wavelength.

Accordingly, the channel vector between the BS and user $k$ is expressed as \cite{15zhu2023modeling}
\begin{equation}
\label{eq: channel vector}
{{\bf{h}}_k}({{\bf{u}}_k}) = {\left({\bf{f}}_k^{\rm{H}}({{\bf{u}}_k}){{\bf{\Sigma }}_k}{{\bf{G}}_k}\right)^{\rm{T}}},
\end{equation}
where ${{\bf{\Sigma }}_k} \in {\mathbb{C}^{L_k^r \times L_k^t}}$ is the path-response matrix (PRM), and the element $[{{\bf{\Sigma }}_k}]_{j,i}$ represents the channel response coefficient between the $i$-th transmit path and the $j$-th receive path for the user $k$, and ${{\bf{G}}_k} = [{{\bf{g}}_k}({{\bf{v}}_1}),{{\bf{g}}_k}({{\bf{v}}_2}),...,{{\bf{g}}_k}({{\bf{v}}_N})] \in {\mathbb{C}^{L_k^t \times {N}}}$ is the transmit field-response matrix (FRM) at the BS. Therefore, even small movements of each MA will change the phase of multi-path channel and the receive FRV, which may result in a significant variation of the channel vector.

\subsection{Adaptive SIC Decoding and Achievable Rate}
The SIC decoding order of NOMA in this paper is defined as the increasing order of the users' channel gains \cite{3zhu2019joint}, \cite{4xiao2018joint}, \cite{7zhao2019secure}. Without loss of generality, let $\pi_k$ denotes the user with the $k$-th highest channel gain, i.e.,
\begin{equation}
\label{eq: channel order}
\left\| {{{\bf{h}}_{\pi_1}}({{\bf{u}}_{\pi_1}})} \right\|_2^2 \ge \left\| {{{\bf{h}}_{\pi_2}}({{\bf{u}}_{\pi_2}})} \right\|_2^2 \ge ... \ge \left\| {{{\bf{h}}_{\pi_K}}({{\bf{u}}_{\pi_K}})} \right\|_2^2.
\end{equation}

For conventional SIC decoding schemes, user $\pi_k$ needs to decode all the signals for other users  with prior decoding order. However, this fixed decoding scheme may result in a significant performance loss because it does not consider the channel correlation among users. For example, if the channel vectors of two users are orthogonal to each other, the optimal strategy is that the BS adopts the maximal ratio transmitting (MRT) precoding vectors and each user directly decodes its signal without interference cancellation. In such a case, the fixed SIC decoding scheme enforces one user to decode the signal for the other user and thus deteriorates the system performance. 

To address this issue, we propose an adaptive decoding scheme in the SIC framework. Specifically, a binary optimization variable $m_{\pi_k,\pi_j}, k \le j$, is used to indicate whether user $\pi_k$ decodes user $\pi_j$'s message. All these binary optimization variables are collected into a matrix called the \textit{decoding indicator matrix}. When $m_{\pi_k,\pi_j}=1$, it implies that user $\pi_k$ needs to decode the message for user $\pi_j$; on the contrary, when $m_{\pi_k,\pi_j}=0$, it means that user $\pi_k$ does not need to decode the message for user $\pi_j$ while treats it as interference. As such, the $K \times K$-dimensional decoding indicator matrix is given by
\begin{equation}
\label{eq: decoding indicator matrix}
\bf{M} = 
\left[ {\begin{array}{*{20}{c}}
{1}&{{m_{{\pi _1},{\pi _2}}}}&{{m_{{\pi _1},{\pi _3}}}}& \cdots &{{m_{{\pi _1},{\pi _K}}}}\\
{0}&{1}&{{m_{{\pi _2},{\pi _3}}}}& \cdots &{{m_{{\pi _2},{\pi _K}}}}\\
{0}&{0}&{1}& \cdots &{{m_{{\pi _3},{\pi _K}}}}\\
 \vdots & \vdots & \vdots & \ddots & \vdots \\
{0}&{0}&{0}& \cdots &{1}
\end{array}} \right].
\end{equation}

% 这一段可以用来解释插图，以及解码图的意义
 It can be seen that the elements in the $k$-th row of the decoding indicator matrix represent the current decoding scheme for user $\pi_k$. In particular, the diagonal elements of the decoding indicator matrix are all ones, i.e., $m_{\pi_k,\pi_k}=1$, $\forall k \in \mathcal{K}$, which ensures that each user decodes its own signal. The lower triangular part of the decoding indicator matrix are all zeros,  because we adopt the increasing-channel-gain order for SIC decoding, where the NOMA user with poor channel conditions does not decode the signals for users with better channel conditions \cite{3zhu2019joint}, \cite{4xiao2018joint}, \cite{14xiao2019user}.
 
Actually, if all elements of the decoding indicator matrix except for the diagonal are zeros, the decoding scheme is degraded into the SDMA, i.e., all other users' signals are treated as interference. If all the elements located in the upper triangular part of the decoding indicator matrix are ones, the decoding scheme becomes the conventional NOMA, i.e., for each user, the signals for users with worse channel conditions are first decoded while those for users with better channel conditions are treated as interference. Thus, the adaptive decoding scheme is a general solution which compromises SDMA and NOMA under different channel conditions of users for improving the system performance when it is optimized in the whole system. 

Then, the received SINR under the adaptive decoding scheme at user $\pi_k$ for decoding its own message can be represented as
\begin{equation}
\label{eq: self decoding SINR}
{\gamma _{{\pi _k} \to {\pi _k}}} = \frac{{{{\left| {{\bf{h}}_{{\pi _k}}^{\rm{H}}({{\bf{u}}_{{\pi _k}}}){{\bf{w}}_{{\pi _k}}}} \right|}^2}}}{{\sum\limits_{i = 1}^K {{{\left| {{\bf{h}}_{{\pi _k}}^{\rm{H}}({{\bf{u}}_{{\pi _k}}}){{\bf{w}}_{{\pi _i}}}} \right|}^2}\left( {1 - {m_{{\pi _{k,}}{\pi _i}}}} \right) + {\sigma ^2}} }}, \forall k \in \mathcal{K}.
\end{equation}
The received SINR under the adaptive decoding scheme at user $\pi_k$ to decode the message of user $\pi_j$ can be expressed as
\begin{equation}
\label{eq: other decoding SINR}
\begin{split}
&{\gamma _{{\pi _k} \to {\pi _j}}}= \\
&\frac{{{{\left| {{\bf{h}}_{{\pi _k}}^{\rm{H}}({{\bf{u}}_{{\pi _k}}}){{\bf{w}}_{{\pi _j}}}} \right|}^2}}}{{\sum\limits_{i = 1}^{j - 1} {{{\left| {{\bf{h}}_{{\pi _k}}^{\rm{H}}({{\bf{u}}_{{\pi _k}}}){{\bf{w}}_{{\pi _i}}}} \right|}^2} + \sum\limits_{i = j}^K {{{\left| {{\bf{h}}_{{\pi _k}}^{\rm{H}}({{\bf{u}}_{{\pi _k}}}){{\bf{w}}_{{\pi _i}}}} \right|}^2}} \left( {1 - {m_{{\pi _{k,}}{\pi _i}}}} \right) + {\sigma ^2}} }}, \\
\end{split}
\end{equation}
with $1 \le k \le K-1$, $k+1 \le j \le K$. Thus, the achievable rate for decoding the message of user $\pi_j$ can be calculated as
\begin{equation}
\label{eq: rate}
\begin{split}   
R_{j}&={\rm log_2}\left(1+\min_{ \{ \pi_k \mid m_{\pi_k, \pi_j} = 1 \}} {\gamma _{\pi_k \to \pi_j}} \right)\\
         & \triangleq {\rm log_2 \left(1+ \gamma _{j}\right)},
\end{split}
\end{equation}
with $1 \le j \le K, 1 \le k \le j$. It is worth noting that, unlike the conventional FPA-NOMA, the achievable rate for each user under the proposed MA-NOMA with adaptive decoding depends on three variables, i.e., the positions of the MAs, the precoding matrix at the BS, and the decoding indicator matrix at users.

\subsection{Problem Formulation}

In this paper, we aim to maximize the minimum achievable rate among multiple users by jointly optimizing the MAs' positions, i.e., ${\bf{\tilde u}}$, the precoding matrix of the BS, i.e., $\bf W$, and the decoding indicator matrix, i.e., $\bf{M}$. Accordingly, the optimization problem can be formulated as
\begin{subequations}\label{eq: prob1}
\begin{align}
{\max_{{\bf{\tilde u}},{\bf{W}},{\bf{M}}}} \: &{\min _{j \in \mathcal{K}}} \: {{R_j}} \label{eq: obj1}\\
{\rm{s}}{\rm{.t.}}\quad &{\rm{tr}}({{\bf{W}}^{\rm{H}}}{\bf{W}}) \le {P_{\max }},\label{eq: cons-power}\\
&{{\bf{u}}_k} \in {{{\cal C}}_k}, \forall k \in {{\cal K}}, \label{eq: cons-move range}\\
&m_{\pi, \pi} \in \{0, 1\}, \label{eq: cons-decoding indicator matrix} 
\end{align}
\end{subequations}
where constraint (\ref{eq: cons-power}) ensures that the total transmit power of the BS does not exceed the maximum budget $P_{\max}$. Constraint (\ref{eq: cons-move range}) indicates that each user's single-MA can only move within the region, $\mathcal{C}_k$. Constraint (\ref{eq: cons-decoding indicator matrix}) indicates that the elements of the decoding indicator matrix are binary variables. It should be emphasized that solving the non-convex problem (\ref{eq: prob1}) is challenging due to the fact that variables $\tilde{\bf{u}}$ and $\bf{W}$ are high-dimensional and highly coupled, which makes it difficult to obtain an optimal solution in polynomial time. For this reason, in the next section, we propose a two-loop iterative optimization algorithm based on HO and AO algorithms to obtain a suboptimal solution for this problem.

\section{Proposed Solution}
Due to the highly-coupled nature of the MA's positions and the precoding matrix, conventional optimization methods may perform poorly and lead to undesired local optima. To overcome this issue, we propose a two-loop iterative algorithm that combines HO method with AO method. In the inner loop, given the positions of MAs, the precoding matrix and the decoding indicator matrix are optimized alternatively. Specifically, we implement SCA to transform the non-convex problem into a convex one and use existing convex optimization tools such as CVXPY in \cite{34diamond2016cvxpy} to optimize the precoding matrix of the BS. Subsequently, a customized greedy search method is employed to optimize the decoding indicator matrix by combining the idea of accepting bad solutions according to a certain probability from simulated annealing (SA) method \cite{35kirkpatrick1983optimization}. In the outer loop, an improved HO algorithm is proposed to update each user's MA position, which can effectively prevent the algorithm from falling into local optima.

\subsection{Precoding Matrix Optimization}
In the inner loop, given the positions of MAs and the decoding indicator matrix, only the precoding matrix $\bf W$ needs to be optimized. The optimization problem (\ref{eq: prob1}) is transformed into:

\begin{subequations}\label{eq: prob W-1}
\begin{align}
{\max_{\bf{W}}}\: &{\min _{j \in \mathcal{K}}} \: {{{\log }_2}\left( {1 + {\gamma _j}} \right)} \label{eq: obj W-1}\\
{\rm{s}}{\rm{.t.}}\:\: &{\rm{tr}} ({{\bf{W}}^{\rm H}}{\bf{W}}) \le {P_{\max }}{\rm{. }} \label{eq: cons-power-W-1}
\end{align}
\end{subequations}

Subproblem (\ref{eq: prob W-1}) is non-convex and we reformulate it into a convex form. Introducing the vector of auxiliary variables ${\bf{q}} = \left[q_1, q_2,...,q_K\right]$, we can express the problem as
\begin{subequations}\label{eq: prob W-2}
\begin{align}
\mathop {\max }\limits_{{\bf{W}},{q_j}} &\mathop {\min }\limits_{j \in \mathcal{K}} {\rm{ }}{q_j}{\rm{ }}\label{eq: obj W-2}\\
{\rm{s}}{\rm{.t.}} \:&{\rm{tr}} ({{\bf{W}}^{\rm H}}{\bf{W}}) \le {P_{\max }}{\rm{, }} \label{eq: cons-power-W-2}\\
&{\gamma _j} \ge {q_j},1 \le j \le K. \label{eq: cons-SINR-q-W-2}
\end{align}
\end{subequations}

\begin{figure*}[!t]
\normalsize
\setcounter{equation}{14} %公式编号的前一个数字，需要手动输入
\begin{subequations}\label{eq: DC SINR>=q}
\begin{align}
\sum\limits_{i = 1}^K {{{\left| {{\bf{h}}_{{\pi _j}}^{\rm{H}}\left( {{{\bf{u}}_{{\pi _j}}}} \right){{\bf{w}}_{{\pi _i}}}} \right|}^2}\left( {1 - {m_{{\pi _j},{\pi _i}}}} \right) + {\sigma ^2}}  & \le \frac{{{{\left| {{\bf{h}}_{{\pi _j}}^{\rm{H}}\left( {{{\bf{u}}_{{\pi _j}}}} \right){{\bf{w}}_{{\pi _j}}}} \right|}^2}}}{{{q_j}}},{\rm{ }}1 \le j \le K, \label{eq: DC SINR>=q k,k}\\
\sum\limits_{i = 1}^{j - 1} {{{\left| {{\bf{h}}_{{\pi _k}}^{\rm{H}}\left( {{{\bf{u}}_{{\pi _k}}}} \right){{\bf{w}}_{{\pi _i}}}} \right|}^2} + \sum\limits_{i = j}^K {{{\left| {{\bf{h}}_{{\pi _k}}^{\rm{H}}\left( {{{\bf{u}}_{{\pi _k}}}} \right){{\bf{w}}_{{\pi _i}}}} \right|}^2}} \left( {1 - {m_{{\pi _k},{\pi _i}}}} \right) + {\sigma ^2}}  & \le \frac{{{{\left| {{\bf{h}}_{{\pi _k}}^{\rm{H}}\left( {{{\bf{u}}_{{\pi _k}}}} \right){{\bf{w}}_{{\pi _j}}}} \right|}^2}}}{{{q_j}}},
1\le k \le K - 1,k + 1 \le j \le K. \label{eq: DC SINR>=q k,j}
\end{align}
\end{subequations}
\hrulefill %加横线
%\vspace*{4pt}//根据自己的需求决定是否使用,垂直间距
\end{figure*}

Although the objective function in problem (\ref{eq: prob W-2}) have been transformed into convex, constraint (\ref{eq: cons-SINR-q-W-2}) is still non-convex, which can be transformed into the difference of convex (DC) functions as (\ref{eq: DC SINR>=q k,k}) and (\ref{eq: DC SINR>=q k,j}). Then, we use

\begin{equation}
\label{eq: taylor expansion}
\begin{split}
&\frac{{{{\left| {{\bf{h}}_{{\pi _k}}^{\rm{H}}\left( {{{\bf{u}}_{{\pi _k}}}} \right){{\bf{w}}_{{\pi _j}}}} \right|}^2}}}{{{q_j}}} \ge
\frac{{2{\mathop{\rm Re}\nolimits} ({\bf{\bar w}}_{\pi_j}^{\rm{H}}{{\bf{h}}_k}\left( {{{\bf{u}}_{{\pi _k}}}} \right){\bf{h}}_k^{\rm{H}}\left( {{{\bf{u}}_{{\pi _k}}}} \right){{\bf{w}}_{\pi_j}})}}{{{{\bar q}_j}}} \\
&-\frac{{{\mathop{\rm Re}\nolimits} ({\bf{\bar w}}_{\pi_j}^{\rm{H}}{{\bf{h}}_k}\left( {{{\bf{u}}_{{\pi _k}}}} \right){\bf{h}}_k^{\rm{H}}\left( {{{\bf{u}}_{{\pi _k}}}} \right){{{\bf{\bar w}}}_{\pi_j}})}}{{{{\bar q}_j}^2}}{q_j}  \buildrel \Delta \over = {{{\cal T}}_{k,j}}{\rm{(}}{{\bf{w}}_{\pi_j}},{q_j},{{{\bf{\bar w}}}_{\pi_j}}{\rm{,}}{{\bar q}_j}{\rm{)}}, \\
\end{split}
\end{equation}
where ${\bf{\bar w}}_{\pi_j}$ and ${\bar q}_j$ denote the values of the variables ${\bf w}_{\pi_j}$ and ${q}_j$ at the $(t-1)$-th iteration, respectively. Thus, constraints (\ref{eq: DC SINR>=q k,k}) and (\ref{eq: DC SINR>=q k,j}) are relaxed into the second-order cone (SOC) form, which are shown as (\ref{eq: SOC SINR k,k}) and (\ref{eq: SOC SINR k,j}), respectively.

% \begin{figure*}[!t]
% \normalsize
% \setcounter{equation}{16} %公式编号的前一个数字，需要手动输入
% \begin{subequations}\label{eq: convex SINR}
% \begin{align}
% {\sum\limits_{i = 1}^K {{{\left| {{\bf{h}}_{{\pi _j}}^{\rm{H}}\left( {{{\bf{u}}_{{\pi _j}}}} \right){{\bf{w}}_{{\pi _i}}}} \right|}^2}\left( {1 - {m_{{\pi _j},{\pi _i}}}} \right) + {\sigma ^2}}}  & \le {{{\cal T}}_{j,j}}\left( {{{\bf{w}}_j},{q_j},{{{\bf{\bar w}}}_j},{{\bar q}_j}} \right), 1 \le j \le K,\label{eq: convex SINR k,k}\\
% {\sum\limits_{i = 1}^{j - 1} {{{\left| {{\bf{h}}_{{\pi _k}}^{\rm{H}}\left( {{{\bf{u}}_{{\pi _k}}}} \right){{\bf{w}}_{{\pi _i}}}} \right|}^2} + \sum\limits_{i = j}^K {{{\left| {{\bf{h}}_{{\pi _k}}^{\rm{H}}\left( {{{\bf{u}}_{{\pi _k}}}} \right){{\bf{w}}_{{\pi _i}}}} \right|}^2}} \left( {1 - {m_{{\pi _k},{\pi _i}}}} \right) + {\sigma ^2}} } & \le {{{\cal T}}_{k,j}}\left( {{{\bf{w}}_j},{q_j},{{{\bf{\bar w}}}_j},{{\bar q}_j}} \right), 1 \le k \le K,k + 1 \le j \le K.\label{eq: convex SINR k,j}
% \end{align}
% \end{subequations}
% \hrulefill %加横线
% %\vspace*{4pt}//根据自己的需求决定是否使用,垂直间距
% \end{figure*}

\begin{figure*}[!t]
\normalsize
\setcounter{equation}{16} %公式编号的前一个数字，需要手动输入
\begin{subequations}\label{eq: SOC SINR}
\begin{align}
&\left\| {{{\left[ {2\sqrt {1 - {m_{{\pi _j},{\pi _1}}}} {\bf{h}}_{{\pi _j}}^{\rm{H}}\left( {{{\bf{u}}_{{\pi _j}}}} \right){{\bf{w}}_{{\pi _1}}},...,2\sqrt {1 - {m_{{\pi _j},{\pi _K}}}} {\bf{h}}_{{\pi _j}}^{\rm{H}}\left( {{{\bf{u}}_{{\pi _j}}}} \right){{\bf{w}}_{{\pi _K}}},2\sigma ,{{{\cal T}}_{j,j}}\left( {{{\bf{w}}_{\pi_j}},{q_j},{{{\bf{\bar w}}}_{\pi_j}},{{\bar q}_j}} \right) - 1} \right]}^{\rm{H}}}} \right\|_2 \notag \\
&\le {{{\cal T}}_{j,j}}\left( {{{\bf{w}}_{\pi_j}},{q_j},{{{\bf{\bar w}}}_{\pi_j}},{{\bar q}_j}} \right) + 1, 1 \le j \le K,\label{eq: SOC SINR k,k}\\
&\left\| {{{\left[ {2 {\bf{h}}_{{\pi _k}}^{\rm{H}}\left({{{\bf{u}}_{{\pi _k}}}} \right){{\bf{w}}_{{\pi _1}}},...,2\sqrt{1-m_{\pi_k,\pi_j}}{\bf{h}}_{{\pi _k}}^{\rm{H}}\left( {{{\bf{u}}_{{\pi _k}}}} \right){{\bf{w}}_{{\pi _j}}},...,2\sqrt{1-m_{\pi_k,\pi_K}}{\bf{h}}_{{\pi _k}}^{\rm{H}}\left( {{{\bf{u}}_{{\pi _k}}}} \right){{\bf{w}}_{{\pi _K}}}, 2\sigma ,{{{\cal T}}_{k,j}}\left( {{{\bf{w}}_{\pi_j}},{q_j},{{{\bf{\bar w}}}_{\pi_j}},{{\bar q}_j}} \right) - 1} \right]}^{\rm{H}}}} \right\|_2  \notag \\
& \le {{{\cal T}}_{k,j}}\left( {{{\bf{w}}_{\pi_j}},{q_j},{{{\bf{\bar w}}}_{\pi_j}},{{\bar q}_j}} \right) + 1, 1 \le k \le K,k + 1 \le j \le K.\label{eq: SOC SINR k,j}
\end{align}
\end{subequations}
\hrulefill %加横线
%\vspace*{4pt}//根据自己的需求决定是否使用,垂直间距
\end{figure*}
% ,...,2{\bf{h}}_{{\pi _k}}^{\rm{H}}\left( {{{\bf{u}}_{{\pi _{k}}}}} \right){{\bf{w}}_{{\pi _{j-1}}}}

With the above derivations, all the constraints are converted to be convex, and the original non-convex problem (\ref{eq: prob W-1}) is transformed into a convex problem as
\begin{subequations}\label{eq: prob convex W}
\begin{align}
\mathop {\max }\limits_{{\bf{W}},{q_j}} &\mathop {\min }\limits_{j \in \mathcal{K}} {\rm{ }}{q_j}{\rm{ }}\label{eq: obj W-2}\\
{\rm{s}}{\rm{.t.}} \:& {\rm{(\ref{eq: SOC SINR k,k})}} \: {\rm and} \:  {\rm {(\ref{eq: SOC SINR k,j})}}, \notag \\
&{\rm{tr}}({{\bf{W}}^{\rm H}}{\bf{W}}) \le {P_{\max }}{\rm{, }} \label{eq: cons-power-convex-W}
\end{align}
\end{subequations}
which can be solved using existing toolboxes such as CVXPY.

\subsection{Decoding Indicator Matrix Optimization}

Given the APV $\bf \tilde u$ and the precoding matrix $\bf W$, we propose a customized greedy search algorithm to optimize the decoding indicator matrix, which can significantly reduce the algorithm complexity. Specifically, we introduce the idea of accepting suboptimal solutions based on a defined probability from the SA algorithm to increase the randomness of the greedy search method, which enables our algorithm to escape from the local optima effectively.

Then, the optimization problem (\ref{eq: prob1}) can be written as a subproblem, which is given by
\begin{subequations}\label{eq: prob M}
\begin{align}
{\max_{{\bf{M}}}} \:&{\min _{j \in \mathcal{K}}} \: {{R_j}} \label{eq: obj2}\\
{\rm{s}}{\rm{.t.}}\:\: &m_{\pi, \pi} \in \{0, 1\}. \label{eq: cons-M-decoding indicator matrix}
\end{align}
\end{subequations}

For simplicity, the minimum achievable rate among all users for given the precoding matrix $\bf W$ and the decoding indicator matrix $\bf M$ can be expressed as
\begin{equation}\label{eq: f=R}
f\left( {{{\bf{W}}}, {{\bf{M}}}} \mid \tilde{\bf{u}} \right) \triangleq{ \mathop {\min }\limits_{j \in \mathcal{K}} {{R_j}} },
\end{equation}
where $R_j$ can be calculated by Eqs. \eqref{eq: self decoding SINR}-\eqref{eq: rate} with given $\bf{W}$ and $\bf{M}$.

The details of our proposed algorithm for customized greedy search is shown in Algorithm \ref{algorithm M}. In line 1, we initialize the decoding indicator matrix and the minimum achievable rate according to the previous inner-loop iteration. In addition, the idea of the SA algorithm is integrated into the greedy search, i.e., accepting bad solutions with a certain probability. Lines 2 to 24 are the procedure of greedy search for optimizing the set of variables of decoding indicator matrix, where each element in this matrix will be visited only once and thus the complexity is significantly reduced.
% Algorithm Greedy
\begin{algorithm}[t]  %其中这里面不能有H不然会报错，不过不影响结果
\label{algorithm M}
	\caption{Greedy search for solving problem (\ref{eq: prob M}).}%算法名字
	\LinesNumbered %要求显示行号
	\KwIn{$K$, $\sigma^2$, ${\bf{H}}\left( {{\bf{\tilde u}}} \right)$, ${{{\bf{W}}^{\left( t \right)}}}$, ${{{\bf{M}}^{\left( t-1 \right)}}}$, ${R^{\left( {t - 1} \right)}}$, $T$, $\xi$.
     } 
    %输入参数
	\KwOut{${\bf{M}}$.} %输出
	Initialize ${\bar{\bf M}}={{\bf M}^{\left(t-1 \right)}}$, $\bar R = {R^{\left( {t - 1} \right)}}$, $R_g = 0$, $R_b = 0$.
    %\;用于换行
    \\
	\For{$k = 1$ {\rm to} $K-1$}
    {   
        \For{$j=k+1$ {\rm to} $K$}
        {
        Update ${\bar{\bf M}}$ according to ${m_{\pi_k,\pi_j}} \leftarrow 1 - {m_{\pi_k,\pi_j}}$. \\
        Calculate the candidate minimum achievable rate ${R_{{\rm{temp}}}}{\rm{ =  }}f\left( {{{\bf{W}}^{^{\left( t \right)}}},{\bar{\bf M}}} \mid \tilde{\bf{u}} \right)$.\\
        \If{${R_{{\rm{temp}}}} > \bar R$}
        {$\bar R \leftarrow {R_{{\rm{temp}}}}.$\\
         $R_g \leftarrow R_{\rm{temp}}$.\\
         ${\bf M}_g \leftarrow \bf{\bar M}$.\\
        }
        \Else
        {
            \If{${rand(0, 1)} < \xi T$}
            {
             \If{$R_{\rm{temp}} > R_b$}
             {
             $R_b \leftarrow R_{\rm{temp}}.$\\
             ${\bf M}_b \leftarrow \bf{\bar M}$.\\
            }
            Recover ${\bar{\bf M}}$ according to ${m_{\pi_k,\pi_j}} \leftarrow 1 - {m_{\pi_k,\pi_j}}$.\\
            }
            \Else
            {
            Recover ${\bar{\bf M}}$ according to ${m_{\pi_k,\pi_j}} \leftarrow 1 - {m_{\pi_k,\pi_j}}$.\\
            }
        }
        }
	}
    Find the maximum candidate minimum achievable rate according to ${R_{\max }} \leftarrow \max \{ R_g, R_b \}$. \\
    Find the corresponding decoding indicator matrix ${\bf{M}}_{\max}$ according to ${R_{\max }}$.\\
	\If{${R_{\max }} > {\rm{ }}{R^{\left( {t - 1} \right)}}$}
    {
    ${\bf{M}} \leftarrow {\bf{M}}_{\max}$.\\
    \Else
    {
        \If{$rand(0, 1) < {{\mathop{\rm e}\nolimits} ^{\frac{{ {{R_{\max }} - {R^{\left( {t - 1} \right)}}} }}{T}}}$}
        {
         ${\bf{M}} \leftarrow {\bf{M}}_{\max}$.
        }
        \Else
        {${\bf{M}} \leftarrow {{\bf M}^{\left( t-1 \right)}}$.}
    }
    }
   
	\KwRet ${\bf M}$
\end{algorithm}

In line 6, if the candidate minimum achievable rate $R_{\rm temp}$ is higher than the previously accepted solution $\bar R$, it indicates a good solution and is accepted directly. Otherwise, the solution is considered a bad solution and its acceptance depends on the probability $\xi T$ (line 12), where $\xi$ is a constant coefficient and $T$ is the temperature. To avoid error accumulation, in lines 12 to 18, we do not update $\bar R$ and $\bar{\bf{M}}$ based on the bad solutions, even though these solutions are accepted. In lines 27-37, we compare the maximum candidate minimum achievable rate, i.e., $R_{\rm max}$, with the minimum achievable rate from the previous inner-loop iteration, i.e., $R^{\left(t-1\right)}$. If $R_{\rm max}$ is less than $R^{\left(t-1\right)}$, we use the Metropolis criterion (line 30) to determine if it will be accepted. Eventually, the algorithm outputs the updated decoding indicator matrix $\bf{M}$.

\subsection{AO Algorithm}

Then, an AO algorithm (i.e., Algorithm \ref{algorithm AO} with inner-loop iterations) is proposed for calculating the max-min achievable rate by alternately optimizing the precoding matrix $\bf W$ and the decoding indicator matrix $\bf{M}$ for a given APV $\bf \tilde u$. In line 1, the decoding indicator matrix is initialized to an identical matrix of size $K \times K$, i.e., SDMA scheme. The temperature $T$ is set to the initial temperature $T_0$ and $\alpha$ is the temperature decay coefficient. At the beginning of the iteration, the temperature is higher, which means the algorithm has a greater probability of accepting suboptimal solutions to increase randomness and avoid local optima. As the iteration progresses and the temperature decreases, the algorithm becomes less likely to accept suboptimal solutions, ensuring convergence. In lines 5 to 13, the precoding matrix $\bf W$ and the decoding indicator matrix $\bf M$ are alternatively optimized. Where $\epsilon_1$ and $\epsilon_2$ are the termination thresholds for the temperature and the change in the minimum achievable rate, respectively. Finally, we can obtain the maximum minimum achievable rate for given APV based on the optimized precoding matrix and decoding indicator matrix. Although the AO algorithm may yield a suboptimal solution, it provides an effective trade-off between performance and complexity for the highly non-convex problems formulated in this paper.
% Algorithm AO
\begin{algorithm}[]  %其中这里面不能有H不然会报错，不过不影响结果
\label{algorithm AO}
    \SetAlgoLined 
	\caption{AO algorithm for optimizing the precoding matrix and the decoding indicator matrix.}%算法名字
	\LinesNumbered %要求显示行号
	\KwIn{$\bf \tilde u$, $K$, $N$, $\sigma^2$, $\lambda$, $\{ {{\bf{\Sigma }}_k}\}$, $\{ {{\bf{G}}_k}\} $, $\{ \theta _{k,j}^r\}$, $\{ \varphi _{k,j}^r\}$, $P_{\rm max}$, $\epsilon_1$, $\epsilon_2$, $\alpha$, $T_0$.
     } 
    %输入参数
	\KwOut{$R\left( {{\bf{\tilde u}}} \right)$, ${\bf W}^{\ast}$, ${\bf M}^{\ast}$.} %输出
	Initialize ${{\bf{M}}^{\left(0\right)}} = {{\bf{I}}_K}$, $T=T_0$, $t=1$.\\
    Randomly initialize ${{\bf{W}}^{\left(0\right)}}$.\\
    Normalize ${{\bf{W}}^{\left(0\right)}} \leftarrow {\sqrt{\frac{P_{\max}}{{\rm tr}\left({{\bf{W}}^{\left(0\right)}}^{\rm H}{{\bf{W}}^{\left(0\right)}}\right)}}{{\bf{W}}^{\left(0\right)}}}$ to satisfy the constraint \eqref{eq: cons-power-convex-W}. \\
    Calculate channel matrix ${\bf{H}}\left( {{\bf{\tilde u}}} \right)$ according to (\ref{eq: channel vector}).\\
    %\\;用于换行
    \While{$T > {\epsilon_1}$} 
    {
    Update ${{\bf{W}}^{\left( t \right)}}$ by solving problem \eqref{eq: prob convex W} for given $ {\bf{H}}\left( {{\bf{\tilde u}}} \right)$ and ${{\bf{M}}^{\left( {t - 1} \right)}}$. \\
    Update ${{\bf{M}}^{\left( t \right)}}$ according to Algorithm \ref{algorithm M} for given ${\bf{H}}\left( {{\bf{\tilde u}}} \right)$ and ${{\bf{W}}^{^{\left( t \right)}}}$.\\
    \If{$\left| {f\left( {{{\bf{W}}^{\left( t \right)}},{{\bf{M}}^{\left( t \right)}}} \mid \tilde{\bf{u}} \right) - f\left( {{{\bf{W}}^{\left( {t - 1} \right)}},{{\bf{M}}^{\left( {t - 1} \right)}}} \mid \tilde{\bf{u}} \right)} \right| < {\epsilon_2}$}
    {
    Break.
    }
    Update $t \leftarrow t+1$.\\
    Update $T=\alpha^t \: T_0$.\\
    }
    Set the precoding matrix ${{\bf W}^{\ast}}$ as ${{\bf{W}}^{\left( t \right)}}$.\\
    Set the decoding indicator matrix ${{\bf M}^{\ast}}$ as ${{\bf{M}}^{\left( t \right)}}$.\\
    Calculate the maximum minimum achievable rate $ R\left( {{\bf{\tilde u}}} \right) = f\left( {{{\bf W}^{\ast}},{{{\bf M}^{\ast}}}} \mid \tilde{\bf{u}} \right)$ for given ${\bf{\tilde u}}$.

	\KwRet $R\left( {{\bf{\tilde u}}} \right)$, ${{\bf W}^{\ast}}$, ${{\bf M}^{\ast}}$.
\end{algorithm}

\subsection{MA Position Optimization}

In the inner loop, the precoding matrix and the decoding indicator matrix are optimized alternatively, while the max-min achievable rate among multiple users is calculated for any given APV, i.e., $ R\left( {{\bf{\tilde u}}} \right) = f\left( {{{\bf W}^{\ast}},{{{\bf M}^{\ast}}}} \mid \tilde{\bf{u}} \right)$ obtained in Algorithm \ref{algorithm AO}. In the outer loop, the positions of MAs are updated such that the original optimization problem (\ref{eq: prob1}) can be transformed as a function of the MAs' positions $\bf \tilde u$, which is given by

\begin{subequations}\label{eq: prob u}
\begin{align}
&\max _{\bf \tilde u} \:\: R\left({{\bf{\tilde u}}} \right)\label{eq: obj u}\\
&{\rm{s.t.}} \: \:  {{{\bf{u}}}_k} \in {{{\cal C}}_k},\forall k \in {{\cal K}}. \label{cons: prob u movable region}
\end{align}
\end{subequations}

To solve the above highly non-convex problem, we introduce a novel intelligent optimization algorithm, i.e., the HO algorithm \cite{36amiri2024hippopotamus}. Unlike the original HO algorithm, we add some tricks to improve the performance of the original HO algorithm. In the HO-based algorithm, each hippopotamus is a candidate solution for the APV, which contains the positions of all the MAs, i.e.,
\begin{equation}\label{hippo vector}
{\bf{\tilde u}}_n^{\left( i \right)} = {[\underbrace {x_{n,1}^{\left( i \right)},y_{n,1}^{\left( i \right)},z_{n,1}^{\left( i \right)}}_{{\rm{MA \:1}}},\underbrace {x_{n,2}^{\left( i \right)},y_{n,2}^{\left( i \right)},z_{n,2}^{\left( i \right)}}_{{\rm{MA\:2}}}...,\underbrace {x_{n,K}^{\left( i \right)},y_{n,K}^{\left( i \right)},z_{n,K}^{\left( i \right)}}_{{\rm{MA}}\:K}]^{\rm{T}}},
\end{equation}
with $1 \le n \le N_H$, $1 \le i \le I_{\rm max}$, where $N_H$ denotes the number of hippopotamuses in the population, and $I_{\rm max}$ denotes the maximum number of iterations for the HO algorithm.

The initialization of the $n$-th hippopotamus's position can be represented as follows
\begin{equation}\label{APV initializing}
{\bf{\tilde u}}_n^{\left( 0 \right)} = b_l + {{\bf{r}}_0} \times \left( {b_u - b_l} \right),1 \le n \le {N_H},
\end{equation}
where $b_l$ and $b_u$ represent the lower and upper bounds on each hippopotamus's movable region, i.e., $-\frac{A}{2}$ and $\frac{A}{2}$, respectively. ${\bf r}_0$ denotes a random vector ranging from 0 to 1.\footnote{For the HO algorithm in this paper, all random numbers, random vectors, and random matrices obey a uniform distribution unless otherwise stated. Additionally, all random vectors have dimension $3K \times 1$.}

In the original HO algorithm, hippopotamuses are directly divided into two parts according to their indices, and their positions are updated in different ways. In this paper, we enhance the original HO algorithm by evenly divided the hippopotamuses into stronger hippopotamuses and weaker hippopotamuses according to their fitness function, i.e., $R\left( {\bf{\tilde u}}\right)$.
Specifically, hippopotamuses with higher fitness values, which are considered as stronger hippopotamuses $\mathcal{S}$, participate in the first phase of position updating. While hippopotamuses with worse fitness values, which are considered as weaker hippopotamuses $\mathcal{W}$, join in the second phase of position updating. All hippopotamuses will participate in the third phase of local position updating.
% i.e., the hippopotamus with the global best fitness) Moreover, the introduction of predator helps to drive weaker hippopotamuses away from the disadvantaged positions

Phase 1 of the HO algorithm is the position update in the river or pond. The stronger hippopotamuses calculate candidate positions and corresponding fitness in two ways, i.e., the calculation of male hippopotamus and the calculation of female hippopotamus. Specifically, the position is updated according to the male hippopotamus's method as \cite{36amiri2024hippopotamus}
\begin{equation}\label{updating position-male hippo}
{\bf{\tilde u}}_n^M = {\bf{\tilde u}}_n^{^{\left( {i - 1} \right)}} + {r_1} \times \left( {{{{\bf{\tilde u}}}^{gbest}} - {{\hat I}_1}{\bf{\tilde u}}_n^{^{\left( {i - 1} \right)}}} \right), n \in {{\cal S}},
\end{equation}
where ${\bf{\tilde u}}_n^M$ is the position calculated by the $n$-th hippopotamus according to the male hippopotamus's method, ${\bf \tilde u}^{gbest}$ is the position of the dominant hippopotamus (i.e., the hippopotamus with the global best fitness), $r_1$ is a random number between 0 and 1. ${{{\hat I}_1}}$ is a random integer selected from 1 and 2.

Next, we define a series of random events as
\begin{equation}
    {{{\cal I}}^F} = \begin{cases}
    {\hat I}_2 \times {{\bf r}_2} + {\varrho},\\
    2 \times {{\bf{r}}_3} - 1,\\
    {{\bf{r}}_4},\\
    {{r_5} \times {{\bf 1}_{3K}}},
\end{cases} \label{eq: event female}\\
\end{equation}
% \begin{align}
% {{{\cal I}}^F} &= \begin{cases}
% {\hat I}_2 \times {{\bf r}_2} + {\varrho},\\
% 2 \times {{\bf{r}}_3} - 1,\\
% {{\bf{r}}_4},\\
% %{{\hat I}_1} \times {{\bf{r}}_4} + \left( {\sim{{\varrho }_2}} \right),\\
% {{{r}}_5},
% \end{cases} \label{eq: event female}\\
% P_I &= \exp \left( { - \frac{i}{{{I_{\max }}}}} \right). \label{decay I}
% \end{align}
where ${\hat I}_2 $ as a random integer selected from 1 and 2, ${\bf r}_j, j=2,3,4$, as random vectors between 0 and 1, $r_5$ denotes a random number between 0 and 1, $\varrho$ as a random integer selected from 0 and 1. ${\cal I}^F$ provides four random vectors with different value ranges, and we can randomly select one vector from them to calculate the position of the hippopotamus. Besides, we define $P_I= {\rm e}^{ - \frac{i}{{{I_{\max }}}}}$ as a probability that decays with iteration. Then, the calculation of position for the $n$-th, $n \in \mathcal{S}$, female hippopotamus is given by \cite{36amiri2024hippopotamus}
\begin{align}
{\bf{\tilde u}}_n^F = \begin{cases}
{\bf{\tilde u}}_n^{^{\left( {i - 1} \right)}} + {{\cal I}}_1^F \odot \left( {{{{\bf{\tilde u}}}^{gbest}} - {{\hat I}_2}{\bf{\tilde u}}_n^{MG}} \right),P_I > 0.6,\\
\Xi, \qquad\qquad\qquad\qquad\quad\quad\quad\quad\:\:\:\:\:\:\:\:{\rm{else,}}
\end{cases} \label{updating position-female hippo}\\
\Xi  = \begin{cases}
{\bf{\tilde u}}_n^{^{\left( {i - 1} \right)}} + {{\cal I}_2^F} \odot \left( {{\bf{\tilde u}}_n^{^{MG}} - {{{\bf{\tilde u}}}^{gbest}}} \right), \:{r_6} > 0.5,\\
\left( b_l + {r_7} \left( {b_u - b_l} \right)\right) \times {{\bf 1}_{3K}},
\quad\quad\:\:\:\: {\rm{else,}}
\end{cases} \label{updating position-female hippo-sub}
\end{align}
where ${\cal I}_1^F$ and ${\cal I}_2^F$ are random vectors randomly drawn from ${\cal I}^F$. ${\bf{\tilde u}}_n^{^{MG}}$ denotes the mean value of the positions for some randomly selected hippopotamuses from all hippopotamuses. $r_6$ and $r_7$ are the random numbers between 0 and 1.

Next, we choose whether or not to update the $n$-th hippopotamus's position based on the fitness of the male hippopotamus and the fitness of the female hippopotamus. In particular, the position update scheme can be expressed as \cite{36amiri2024hippopotamus}
\begin{equation}
{\bf{\tilde u}}_n^{^{\left( i \right)}} = \begin{cases}
{\bf{\tilde u}}_n^{^M}, {\text {if }} {R\left( {{\bf{\tilde u}}_n^{^M}} \right)} > \max \left\{ R\left( {{\bf{\tilde u}}_n^{^{\left( {i - 1} \right)}}} \right), R\left( {{\bf{\tilde u}}_n^{^{F}}} \right) \right\},\\
{\bf{\tilde u}}_n^{^F}, {\text{ if }} {R\left( {{\bf{\tilde u}}_n^{^F}} \right)} > \max \left\{ R\left( {{\bf{\tilde u}}_n^{^{\left( {i - 1} \right)}}} \right), R\left( {{\bf{\tilde u}}_n^{^{M}}} \right) \right\},\\
{\bf{\tilde u}}_n^{^{\left( {i - 1} \right)}}, {\text{ otherwise,}}
\end{cases}\label{update position-stronger}
\end{equation}
with $n \in {\mathcal{S}}$. We can see that the update will only be executed if a better position is searched.

Phase 2 of the HO algorithm is hippopotamuses defense against predators. The weaker hippopotamuses participate in this update, and a predator is introduced to guide the position update of the weaker hippopotamuses. The predator generation method is given by \cite{36amiri2024hippopotamus}  
\begin{align}
{{\bf{\tilde u}}^P} &= b_l + {{\bf{r}}_8} \times \left( {b_u - b_l} \right),\label{position of predator}\\
{{\bf d}_n} & = \left| {{{{\bf{\tilde u}}}^P} - {\bf{\tilde u}}_n^{\left( i-1 \right)}} \right|,n \in {{\cal W}},\label{distance-predator-hippo}
\end{align}
where ${{\bf{\tilde u}}^P}$ denotes the position of the predator and ${\bf{r}}_8$ is a random vector between 0 and 1. Eq. (\ref{distance-predator-hippo}) represents the element-wise distance between the $n$-th hippopotamus and the predator. The position updating scheme for the weaker hippopotamuses when facing a predator can be represented as \cite{36amiri2024hippopotamus}
\begin{equation}
{\bf{\tilde u}}_n^W = \begin{cases}
{\bf{r}}_n^{levy} \odot {{{\bf{\tilde u}}}^P} + {\frac{{{r_b}}}{{{r_c} - {r_d}\cos \left( {2\pi {r_g}} \right)}}} \times {\frac{1}{{{\bf d}_n}}},\\
\qquad\qquad\qquad\qquad\:\: {\rm{ }}R\left( {{{{\bf{\tilde u}}}^P}} \right) > R\left( {{\bf{\tilde u}}_n^{^{\left( {i - 1} \right)}}} \right),\\
{\bf{r}}_n^{levy} \odot {{{\bf{\tilde u}}}^P} +  {\frac{{{r_b}}}{{{r_c} - {r_d}\cos \left( {2\pi {r_g}} \right)}}} \times {\frac{1}{{2{{\bf d}_n} + {{\bf{r}}_9}}}},\\
\qquad\qquad\qquad\qquad\:\: R\left( {{{{\bf{\tilde u}}}^P}} \right) \le R\left( {{\bf{\tilde u}}_n^{^{\left( {i - 1} \right)}}} \right),
\end{cases} \label{update position-weaker}    
\end{equation}
with $n \in {\mathcal{W}}$, where ${\bf{r}}_n^{levy}$ denotes a random vector following Levy distribution. Specifically, a matrix follows the Levy distribution (of size $3K \times N_H$) can be obtained by Eqs. (\ref{levy function}) and (\ref{sigma u}), from which the elements in the $n$-th column are taken as ${\bf{r}}_n^{levy}$. Furthermore, $r_b$ is a random number between 2 and 4, $r_c$ is a random number between 1 and 1.5, $r_d$ is a random number between 2 and 3, $r_g$ is a random number between -1 and 1, ${\bf{r}}_9$ is a random vector between 0 and 1. $\frac{1}{\bf d}_{n}$ represents the vector with element-wise reciprocal entries of ${\bf d}_{n}$.

The mathematical model for Levy's random movement is denoted as 

\begin{align}
{{\bf{R}}^{levy}}\left( \beta  \right) = 0.05 \times \frac{{{{\bf{R}}_u} \times {\sigma _u}}}{{{{\left| {{{\bf{R}}_v}} \right|}^{\frac{1}{\beta }}}}}, \label{levy function}\\
{\sigma _u} = {\left[ {\frac{{\Gamma \left( {1 + \beta } \right)  \sin \left( {\frac{{\pi \beta }}{2}} \right)}}{{\Gamma \left( {\frac{{1 + \beta }}{2}} \right)\beta \times {2^{\frac{{\left( {\beta  - 1} \right)}}{2}}}}}} \right]^{\frac{1}{\beta }}}, \label{sigma u}
\end{align}
where $\beta$ is a constant, ${\bf R}_u$ and ${\bf R}_v$ are random matrices between 0 and 1, both of which have dimensions $3K \times N_H$. $\left|{\bf R}_{v}\right|$ represents the element-wise absolute value of ${\bf R}_{v}$. The calculation of $\sigma_u$ is presented in Eq. (\ref{sigma u}), where $\Gamma$ is an abbreviation for the Gamma function.
 
Then, the position update method for the weaker hippopotamus is represented as \cite{36amiri2024hippopotamus}, for $n \in {{\cal W}},$
\begin{equation}
{\bf{\tilde u}}_n^{^{\left( i \right)}} = \begin{cases}
{\bf{\tilde u}}_n^{^W},& R\left( {{\bf{\tilde u}}_n^{^W}} \right) > R\left( {{\bf{\tilde u}}_n^{^{\left( {i - 1} \right)}}} \right),\\
{\bf{\tilde u}}_n^{^{\left( i-1 \right)}},&{\rm{else.}}
\end{cases} \label{update position-predator-fitness}
\end{equation}

In Phase 2, the weaker hippopotamuses with poorer fitness means that they are at the disadvantaged positions. In nature, the weaker hippopotamuses are often far away from the herd and more likely to encounter predators. For this reason, the predators are introduced to guide the movement of weaker hippopotamuses, which drives the weaker hippopotamuses to jump out of the local optima efficiently.

Phase 3 is hippopotamuses escaping from predators. All hippopotamuses perform a local search. This idea is inspired by the fact that hippopotamuses move to the nearest river or pond when encountering the predator, i.e., searching for the optimal position in the neighborhood, which helps improve the local search capability of the HO algorithm. To simulate this behavior, an effective way is to identify a potential safe zone (river or pond) in the vicinity of hippopotamus and decide whether to update its position by comparing the fitness of the hippopotamus in the safe zone with that of hippopotamus in its current position. The position calculation method for searching the local safe zone can be expressed as \cite{36amiri2024hippopotamus}
\begin{equation}
{\bf \tilde u}_n^{local} = {\bf \tilde u}_n^{\left( i \right)} + {r_{10}} \times \left( {b}_l^{local} + {\mathcal I}_1^{local} \left( {b}_u^{local} - {b}_l^{local} \right) \right),\label{update position-local}    
\end{equation}
with $1 \le n \le N_H$, where ${b_l^{local}}=\frac{b_l}{i}$ and ${b_u^{local}}=\frac{b_u}{i}$ are the lower and upper bounds for local search, respectively. It can be seen that the range of the local search shrinks with iterations, ensuring the stability and convergence of the HO algorithm. $r_{10}$ and $r_{12}$ denote random numbers between 0 and 1, ${\bf r}_{11}$ is a random vector between 0 and 1, $r_{13}$ is a random number that follows a standard normal distribution, and ${{\cal I}}_1^{local}$ is a random vector can be obtained by randomly selecting from the event ${{\cal I}}^{local}$, which is given by

% \begin{align}
% l{b^{local}} &= \frac{{lb}}{t}, \label{lower bound decay}\\
% u{b^{local}} &= \frac{{ub}}{t}, \label{upper bound decay}\\
% {{{\cal I}}^{local}} &= \begin{cases}
% 2 \times {{\bf{r}}_{11}} - 1,\\
% {r_{12}},\\
% {r_{13}}.
% \end{cases} \label{event-local}
% \end{align}

\begin{align}
{{{\cal I}}^{local}} &= \begin{cases}
2 \times {{\bf{r}}_{11}} - 1,\\
{r_{12} \times {\bf 1}_{3K}},\\
{r_{13} \times {\bf 1}_{3K}}.
\end{cases} \label{event-local}
\end{align}

Then, the position update strategy for local search is represented as, for $ 1 \le n \le {N_H},$
\begin{equation}
{\bf{\tilde u}}_n^{^{\left( i \right)}} = \begin{cases}
{\bf{\tilde u}}_n^{local},& R\left( {{\bf{\tilde u}}_n^{local}} \right) > R\left( {{\bf{\tilde u}}_n^{^{\left( i \right)}}} \right),\\
{\bf{\tilde u}}_n^{^{\left( i \right)}},&{\rm{   else}}.
\end{cases}\label{update position-local-fitness}\\
\end{equation}

The details of the improved HO algorithm are shown in Algorithm \ref{algorithm u}.\footnote{Since a fixed decoding order is adopted in this paper, i.e., Eq. (\ref{eq: channel order}), it is necessary to reorder the users by updating ${\pi}_{k}$ according to the user's channel quality when the MAs' positions are updated in Algorithm \ref{algorithm u}.} To guarantee that each MA moves within the feasible region, i.e., to satisfy the constraint (\ref{cons: prob u movable region}), we use the mapping function ${{\cal B}\left({\bf \tilde u}\right)}$ to project the positions of the MAs beyond the bounds to the nearest bounds, i.e.,
% In the HO algorithm, numerous random variables are introduced to enhance the randomness of the algorithm, which is very effective for complex problems that tend to fall into local optima.

\begin{equation}
\label{eq: movable region-map function}
{\left[ {{{\cal B}}({\bf{\tilde u}})} \right]_i} = \begin{cases}
- \frac{A}{2}, &{\text{if }} {[{\bf{\tilde u}}]_i} <  - \frac{A}{2}, \\ 
 \frac{A}{2}, &{\text{if }} {[{\bf{\tilde u}}]_i} >  \frac{A}{2}, \\{[{\bf{\tilde u}}]_i}, &{\text{otherwise.}} 
\end{cases}
\end{equation}

\begin{algorithm}[t]  %其中这里面不能有H不然会报错，不过不影响结果
\label{algorithm u}
	\caption{The improved HO algorithm for solving problem \eqref{eq: prob1}.}%算法名字
	\LinesNumbered %要求显示行号
	\KwIn{$\bf \tilde u$, $K$, $N$, $\sigma^2$, $\lambda$, $\{ {{\bf{\Sigma }}_k}\}$, $\{ {{\bf{G}}_k}\} $, \{$\mathcal{C}_k\}$, $\{ \theta _{k,j}^r\}$, $\{ \varphi _{k,j}^r\}$, $P_{\rm max}$, $\epsilon_1$, $\epsilon_2$, $\xi$, $\alpha$, $T_0$, $N_H$, $\beta$, $I_{\max}$.
     } 
    %输入参数
	\KwOut{${{\bf{\tilde u}}}$, $\bf W$, $\bf{M}$.} %输出
	Initialize the positions for $N_H$ hippos based on (\ref{APV initializing}).\\
    Calculate the fitness for each hippo using Algorithm \ref{algorithm AO}.\\
    
    %\\;用于换行
    \For{$i=1$ \rm to $I_{\max}$}
    {   
        Divide the hippos into stronger hippos $\cal{S}$ and weaker hippos $\cal{W}$ according to their fitness.\\
        Update the position of the dominant hippo ${{{{\bf{\tilde u}}}^{gbest}}}$.\\
        % $\textbf{Phase 1}$: The position update in the river or pond.\\
        \For{ \rm each $n \in \cal{S}$}
        {
        Calculate the position for the $n$-th hippo using (\ref{updating position-male hippo}), (\ref{updating position-female hippo}) and \eqref{eq: movable region-map function}.\\
        Evaluate the fitness for the $n$-th hippo using Algorithm \ref{algorithm AO} and update position using (\ref{update position-stronger}).\\
        }
        %$\textbf{Phase 2}$:  Hippopotamuses defense against predators.\\
        \For{\rm each $n \in \cal{W}$}
        {
        Generate the predator ${{\bf \tilde u}^P}$ using (\ref{position of predator}).\\
        Calculate the position for the $n$-th hippo using (\ref{update position-weaker}) and \eqref{eq: movable region-map function}.\\
        Evaluate the fitness for the $n$-th hippo using Algorithm \ref{algorithm AO} and update position using (\ref{update position-predator-fitness}).\\
        }
        %$\textbf{Phase 3}$:  Hippopotamuses escaping from predators.\\
        Update the local bounds according to $b_l^{local}=\frac{b_l}{i}$ and $b_u^{local}=\frac{b_u}{i}$, respectively.\\
        \For{$n=1$ \rm to $N_H$}
        {
        Calculate the position for the $n$-th hippo using (\ref{update position-local}) and \eqref{eq: movable region-map function}.\\
        Evaluate the fitness for the $n$-th hippo using Algorithm \ref{algorithm AO} and update position using (\ref{update position-local-fitness}).\\
        }
    }
    Obtain the suboptimal APV by ${\bf \tilde u}={{{{\bf{\tilde u}}}^{gbest}}}$.\\
    Calculate the corresponding the precoding matrix $\bf W$ and the decoding indicator matrix $\bf{M}$ according to Algorithm \ref{algorithm AO}.\\
	\KwRet ${{\bf{\tilde u}}}$, $\bf W$, $\bf{M}$.
\end{algorithm}

\subsection{Computational Complexity}
In the inner loop, the complexity mainly comes from solving problem (\ref{eq: prob convex W}), problem (\ref{eq: prob M}), and Algorithm \ref{algorithm AO}. Since problem (\ref{eq: prob convex W}) can be converted to a second-order cone programming (SOCP), solving it has a complexity of ${\mathcal{O}}\left( K^3N\right)$, which depends on the number of optimization variables, the number of constraints, and the dimensions of the constraints\cite{37lobo1998applications}. Thus, Algorithm \ref{algorithm M} for solving problem (\ref{eq: prob M}) has a complexity of $\mathcal{O}\left(K^4N\right)$. In Algorithm \ref{algorithm AO}, the termination condition is determined by the temperature decay coefficient $\alpha$, the initial temperature $T_0$, and the threshold $\epsilon_1$, so the total complexity of the inner loop is $\mathcal{O}\left( K^4N{\log _\alpha }\frac{\epsilon_1}{{{T_0}}}\right)$. The complexity of the outer loop is determined by the number of hippopotamuses $N_H$ and the maximum number of iterations $I_{\max}$. Thus, the total computational complexity for solving problem \eqref{eq: prob1} using Algorithm \ref{algorithm u} is given by $\mathcal{O}\left({I_{\max}}{N_H}{K^4N}{ {\log _\alpha }\frac{\epsilon_1}{{{T_0}}}}\right)$.

\section{Simulation Results}

In this section, we show the simulation setup as well as the simulation results to evaluate the performance of our proposed MA-aided NOMA communication system. 

\subsection{Simulation Setup}
In the simulations, user $k$ is randomly generated around the BS and the distance between them follows a uniform distribution, i.e., $d_k \sim {\mathcal{U}}{\left[{20, 100}\right]}$. A geometric channel is applied, which implies that the numbers of transmit and receive paths are the same for each user, i.e., $L_k^r = L_k^t = L, 1 \le k \le K$. For user $k$, the values of the diagonal elements of its PRM follow a CSCG distribution, i.e., ${\cal{CN}}{\left(0,{c_k^2/L}\right)}$, where ${c_k^2}={g_0}{d_k^{-\zeta}}$ is the expected value of the channel gain for user $k$, $g_0$ denotes the expected value of the average channel gain at a reference distance of 1 m, and $\zeta$ is the path loss exponent. It is important to note that to avoid unfair comparison of simulation results, the variance of the CSCG distribution of the PRM at user $k$ is normalized over the number of paths to ensure that the average channel gain is the same for different numbers of paths. The elevation and azimuth AoAs are obtained with same probability for each user in all directions within the front half-space of its antenna, which depends on the model in \cite{15zhu2023modeling}, i.e., the user's elevation and azimuth AoAs follow a joint probability density function (PDF), ${f_{{\rm{AoA}}}}\left( {\theta _{k,l}^r,\phi _{k,l}^r} \right) = \frac{{\cos \theta _{k,l}^r}}{{2\pi }}$, $\theta _{k,l}^r,\phi _{k,l}^r \in \left[ { - \pi /2,\pi /2} \right]$, $1 \le k \le K$, $1 \le l \le L$. Similarly, the elevation and azimuth AoDs at the BS obey the PDF similar to AoAs. In the simulation, the MAs can move flexibly in a limited 3D space, i.e., $\left[ { - A/2,A/2} \right] \times \left[ { - A/2,A/2} \right] \times \left[ { - A/2,A/2} \right]$. The remaining simulation parameters are shown in Table I.

\subsection{Convergence Performance of Proposed Algorithms}

\begin{figure}[h]
\centering
\includegraphics[width=3in]{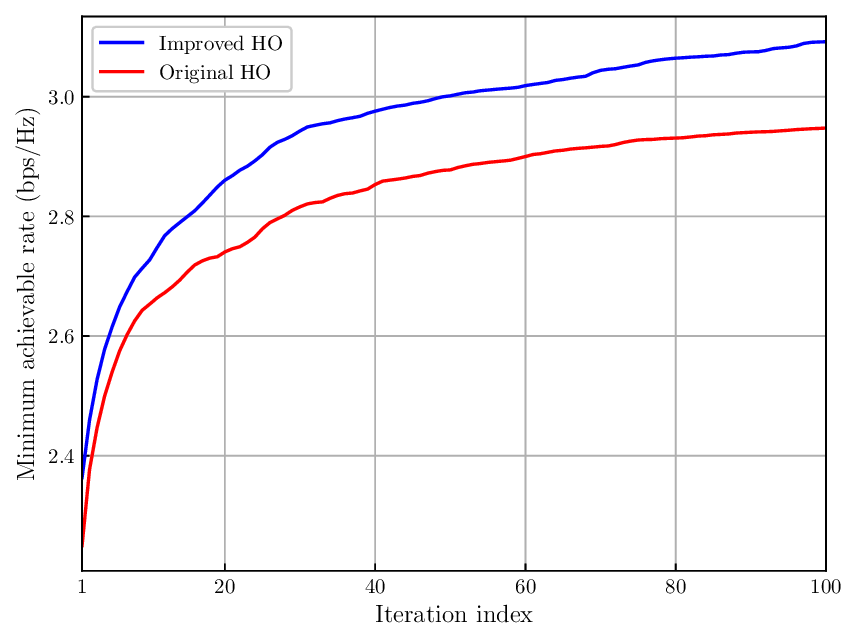}
\caption{Evaluation of the convergence of the proposed improved HO algorithm and the original HO algorithm.}
\label{fig: convergence of algorithm}
\end{figure}

Fig.~\ref{fig: convergence of algorithm} illustrates the convergence of the proposed improved HO algorithm for the considered MA-aided NOMA communication system. In addition, to verify the effectiveness of our improved HO algorithm, we also present the convergence of the original HO algorithm in the same simulation settings. For both algorithms, the minimum achievable rate among all users increases with the iteration index and stabilizes after approximately 80 iterations (less than 0.05 bps/Hz), which indicates a fast convergence performance. Moreover, for the improved HO algorithm, the minimum achievable rate is increased from 2.35 bps/Hz to 3.09 bps/Hz, improving the rate performance by 31.5\%. 

\begin{table}[!h]
\label{Table 1}
\begin{center}
\caption{Simulation Parameters}
\label{tab1}
\begin{tabular}{| c | c | c |} % 三个属性里的数据陈列方式
\hline
Parameter & Description & Value\\
\hline
$N$& Number of antennas at the BS& 4\\
\hline
$K$& Number of users & 6\\ 
\hline
$L$& Number of channel paths for each user & 10\\
\hline 
$\lambda$& Carrier wavelength& 0.01 m\\
\hline 
$g_0$& Average channel power gain at reference distance& -40 dB\\
\hline 
$\zeta$& Exponent of path loss& 2.8\\
\hline 
${\sigma}^2$& Noise power& -80 dBm\\
\hline 
$A$& Length of the sides of movable region ${\cal{C}}_k$& $2\lambda$\\
\hline 
$P_{\max}$& Maximum transmit power for the BS& 30 dBm\\
\hline 
$\xi$& Probability parameters in Algorithm \ref{algorithm M}& 0.1\\
\hline 
$\alpha$&Temperature decay coefficient & 0.8\\
\hline 
$T_0$&Initial temperature & 5\\
\hline 
$\epsilon_1$& Temperature threshold for terminating Algorithm \ref{algorithm AO}& $10^{-3}$\\
\hline 
$\epsilon_2$& Increment threshold for terminating Algorithm \ref{algorithm AO}& $10^{-3}$\\
\hline 
$N_H$& Number of hippopotamuses& 50\\
\hline 
$\beta$& Constant of levy distribution& 1.5\\
\hline 
$I_{\max}$& Maximum number of iterations in Algorithm \ref{algorithm u}& 100\\
\hline 
\end{tabular}
\end{center}
\end{table}

\subsection{Performance Comparison with Benchmark Schemes}
The proposed algorithm is labeled as ``MA-NOMA''. In addition, five benchmark schemes are defined as ``MA-NOMA with fixed SIC'', ``MCP-NOMA'', ``FPA-NOMA'', ``MA-SDMA", and ``FPA-SDMA" for performance comparison. For all six schemes, the BS is equipped with fixed-position UPA with $N$ antennas.

\begin{enumerate}
    \item 
    \textbf{MA-NOMA:} This scheme is our modeled MA-aided downlink NOMA communication with adaptive SIC decoding, where each user is equipped with a single-MA.
    \item 
    \textbf{MA-NOMA with fixed SIC:} This scheme is MA-aided NOMA communication with fixed SIC decoding, i.e., the conventional NOMA decoding scheme. We jointly optimize the antennas' positions and the precoding matrix according to Algorithm \ref{algorithm u}.
    \item 
    \textbf{MCP-NOMA:} This scheme is an application of the maximum channel power (MCP) method under NOMA with adaptive SIC decoding. The antenna of each user is placed at the position where the channel power gain is maximized. Then, Algorithm \ref{algorithm AO} is used to jointly optimize the precoding matrix and the decoding indicator matrix.
    \item 
     \textbf{FPA-NOMA:} This scheme is the NOMA with adaptive SIC decoding while each user is equipped with a single-FPA. Specifically, we place each user's antenna at the origin of the movable region.
    \item 
    \textbf{MA-SDMA:} In this scheme, each user is equipped with a single-MA, but the multiple access method is SDMA. In the outer loop, we optimize the antennas' positions using Algorithm \ref{algorithm u}. In the inner loop, we use the ZF-based method proposed in \cite{30pi2023multiuser} to optimize the precoding matrix.
    \item 
    \textbf{FPA-SDMA:} This scheme is conventional SDMA. The users are equipped with a single-FPA and we use the ZF-based approach proposed in \cite{30pi2023multiuser} to optimize the precoding matrix.
    
\end{enumerate}

\begin{figure*}[t]
    \centering
    \begin{minipage}{0.49\linewidth}
        \centering
        \includegraphics[width=3in]{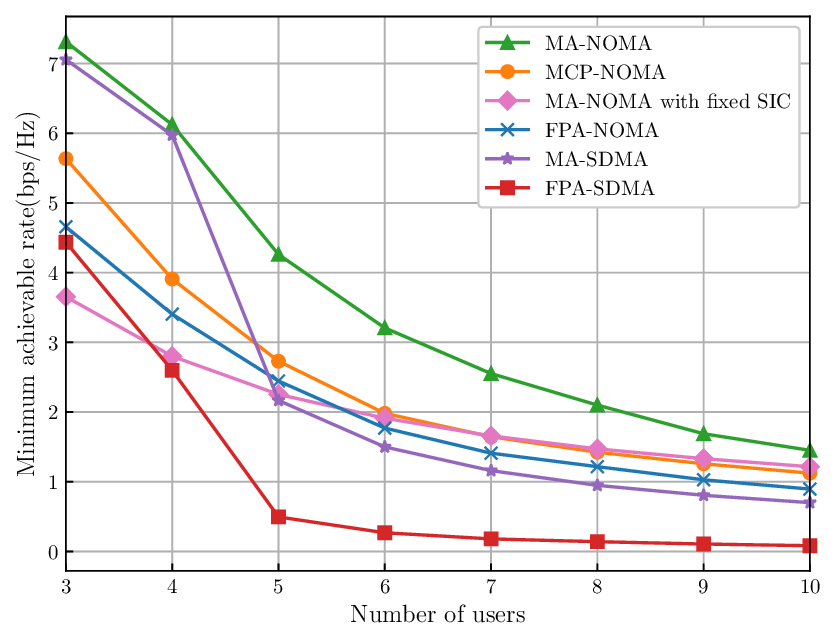}
        \caption{Minimum achievable rates for different schemes versus number of users.}
        \label{user vs rate}
    \end{minipage}
    \hfill
    \begin{minipage}{0.49\linewidth}
        \centering
        \includegraphics[width=3in]{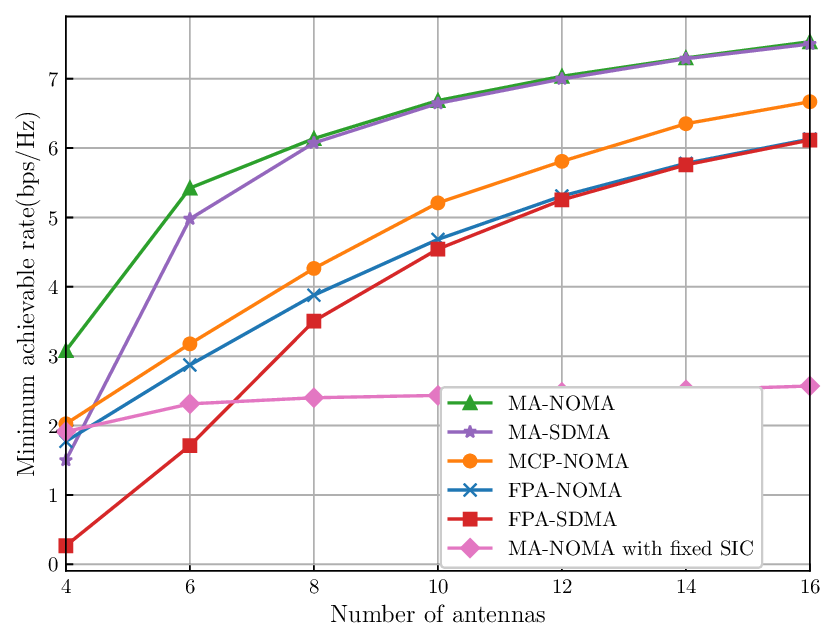}
        \caption{Minimum achievable rates for different schemes versus number of transmit antennas.}
        \label{Tx vs rate}
    \end{minipage}
\end{figure*}

\begin{figure*}[t]
    \centering
    
    \begin{minipage}{0.49\linewidth}
        \centering
        \includegraphics[width=3in]{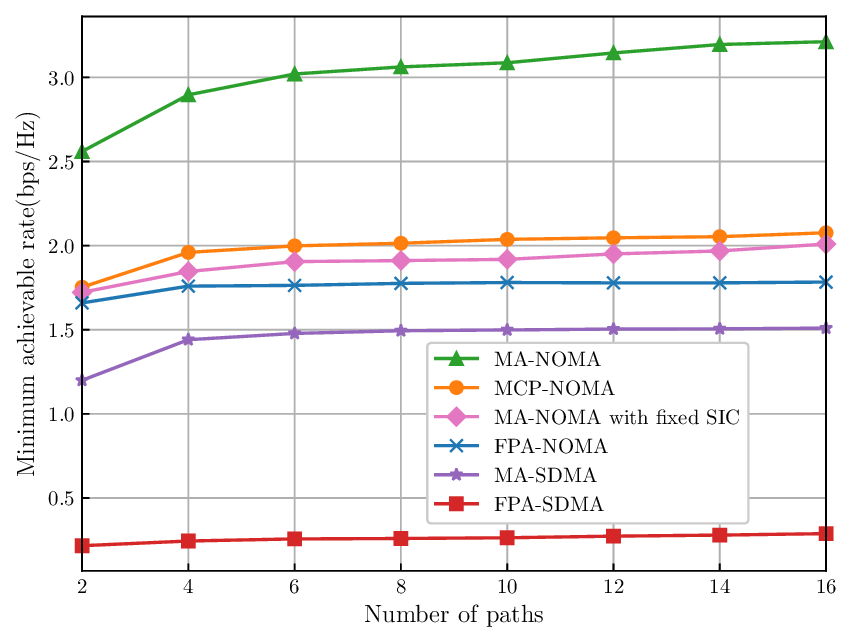}
        \caption{Minimum achievable rates for different schemes versus number of paths.}
        \label{path vs rate}
    \end{minipage}
    \hfill
    \begin{minipage}{0.49\linewidth}
        \centering
        \includegraphics[width=3in]{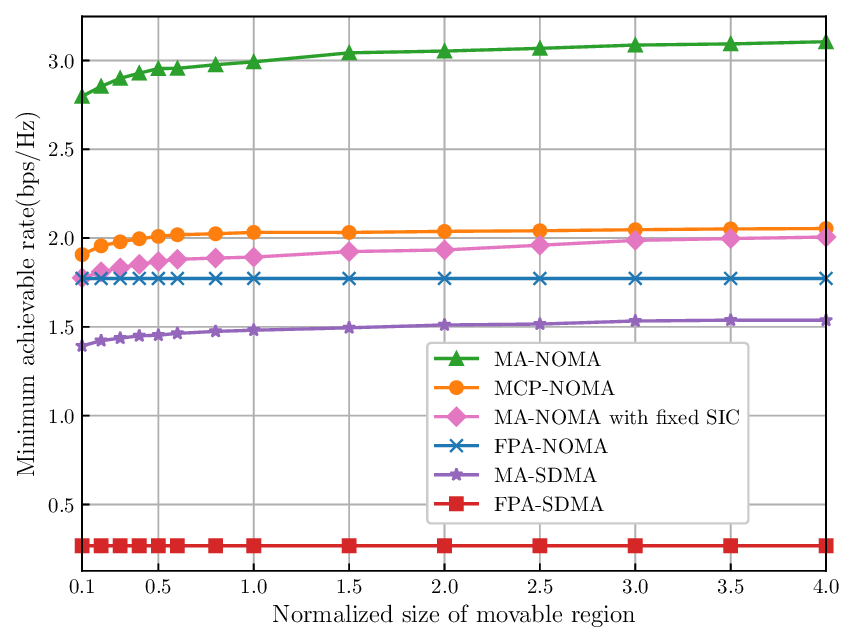}
        \caption{Minimum achievable rates for different schemes versus normalized size of the movable region.}
        \label{region vs rate}
    
    \end{minipage}
\end{figure*}

Fig. \ref{user vs rate} compares the minimum achievable rate versus the number of users for different schemes. The proposed scheme outperforms all other benchmark schemes. Further, the minimum achievable rate of all schemes decreases with the increased number of users. The reason is that the increase in the number of users leads to an increase in inter-user interference, especially when $K > N$.
%i.e., the number of users is more than the number of antennas.
In particular, both SDMA schemes suffer from a devastating performance degradation because it is difficult for the transmit antennas to utilize the spatial DoFs of the channel to calculate appropriate beamforming vectors to reduce inter-user interference. %In other words, ZF-based interference cancellation is unable to operate effectively. 
However, the NOMA-based schemes reveal advantageous performance in multi-user communication.
% when the number of users is more than the number of antennas, the advantage of NOMA is revealed. 
% All four schemes employing NOMA yield better performance than SDMA, even though the MA-SDMA scheme employing MAs still performs worse than any of the NOMA schemes. 
% which proves that NOMA offers a great performance advantage in multi-user communication.
% Moreover, when $K \le N$, the two NOMA schemes without MA-aided that utilize the adaptive SIC decoding, i.e., MCP-NOMA and FPA-NOMA, still outperform FPA-SDMA. However, both of them have lower performance than MA-SDMA, which indicates that MA can provide a great performance boost. 
% It is worth noting that when $K$ is small, the difference between our proposed MA-NOMA and MA-SDMA is very narrow because both effectively eliminate inter-user interference. In this case, the NOMA decoding scheme approaches that of SDMA, meaning that each user under NOMA does not need to decode the signals of many other users. When $K$ is large, the minimum achievable rate of MA-NOMA is only slightly higher than that of MA-NOMA with fixed SIC, which is the fact that the conventional NOMA decoding scheme is suitable for handling simultaneous communication among the excess number of users. 
Notably, as the number of users increases, the performance of the MA-NOMA with fixed SIC scheme rises from the worst case to the second best case, which proves the importance of applying our proposed flexible decoding scheme.

In Fig. \ref{Tx vs rate}, we compare the minimum achievable rate versus the number of antennas for different schemes. 
%It can be seen that the proposed scheme outperforms all other benchmark schemes.
As the number of antennas increases, the minimum achievable rate also increases because the spatial multiplexing and beamforming gains increase.
% For the SDMA schemes, the antennas can better utilize the spatial DoFs, allowing the better interference cancellation and thus enhancing the beamforming gain. 
% However, there is an upper bound to the achievable rate increase from the number of antennas due to the constraint on the maximum transmit power. 
% When $N=4$, the number of antennas is less than the number of users, and the NOMA-based schemes are more dominant than the SDMA-based schemes. 
% As the number of antennas increases, the ZF-based SDMA schemes get good interference cancellation and hence their performance has improved dramatically. 
Thanks to the adaptive decoding scheme, the NOMA-based schemes achieve a better performance than SDMA when the number of antennas is more than the number of users.
% Similarly, the MA-equipped schemes demonstrate their advantages, enabling MA-SDMA to exhibit performance second only to MA-NOMA. 
% It is worth noting that as the number of antennas increases, the performance of both MA-NOMA/MA-SDMA and FPA-NOMA/FPA-SDMA converges to their respective same upper bounds, i.e., the situation where the interference is almost completely eliminated, and the same suboptimal antenna positions are found. 
% However, MA-NOMA with fixed SIC scheme obtains poor performance as the number of antennas increases. Its performance lags far behind the SDMA-based strategies as well as other NOMA with adaptive SIC strategies, which suggests that the fixed SIC decoding scheme does not effectively exploit antenna gain to mitigate inter-user channel correlation, leading to performance degradation.
However, MA-NOMA with fixed SIC scheme obtains the worst performance, which suggests that the fixed SIC decoding scheme does not effectively exploit channel correlation to mitigate inter-user interference, leading to performance degradation.
% From Fig. \ref{user vs rate} and Fig. \ref{Tx vs rate}, we can conclude that considerably improving the performance of the system in our scenario includes three main aspects: placing the antennas in the suitable positions, using adaptive SIC decoding and increasing the number of antennas. All three methods can enhance the channel gain or reduce inter-user interference.

In Fig. \ref{path vs rate}, we compare the minimum achievable rate versus the number of channel paths for different schemes. Our proposed scheme still outperforms all other schemes. 
Moreover, the minimum achievable rate of all schemes increases with the number of channel paths. This is because higher multi-path diversity gain and lower inter-channel correlation are yielded as the number of paths increases. The introduction of MAs further improves the performance as MAs’ positions can be adjusted to reduce the channel correlation between different users.
Therefore, both MA-aided schemes attain obvious performance enhancements, while the performance of FPA-NOMA and FPA-SDMA grows very limited with the increase in the number of paths. 
% On the one hand, both FPA-NOMA and FPA-SDMA are unable to find more suitable positions by moving their antennas. On the other hand, in FPA-SDMA, due to $ N \le K$, it is challenging to mitigate the inter-channel correlation by using beamforming vectors. Thus, even if the number of paths increases, the rate improvement is very limited. As the number of paths increases further, the change in the minimum achievable rate becomes negligible. This is because the periodicity of the channel gain deteriorates as the number of paths increases, which requires a larger movable region to identify the appropriate spatial locations.

Fig. \ref{region vs rate} compares the minimum achievable rate of different schemes versus the size of the normalized movable region of MAs at users, where the size of the movable region is normalized by the carrier wavelength (i.e., $A/\lambda$ ). 
% Our proposed MA-NOMA scheme maintains a large lead over the other schemes in performance.
Both the MA-based schemes and MCP-NOMA achieve an increase in the minimum achievable rate as the normalized movable region increases. This is because as the size of the movable region increases, more channel features can be exploited, e.g., the antennas can be placed to the position with better channel quality or less inter-user interference. 
% However, further increasing the normalized movable region does not result in a significant increase in the minimum achievable rate due to the optimal channel gain point is already contained within the smaller region. 
% Owing to the characteristic of the CSCG distribution, the periodicity of the channel gain is insignificant with a higher number of paths, and thus a larger movable region is also needed to find the optimal antenna deployment position. 
% In our simulation setup, the suboptimal positions of MAs can be found within a smaller normalized region, which also implies that MAs can be deployed in some small space-constrained devices.

\begin{figure*}[t]
    \centering
    \begin{minipage}{0.49\linewidth}
        \centering
        \includegraphics[width=3in]{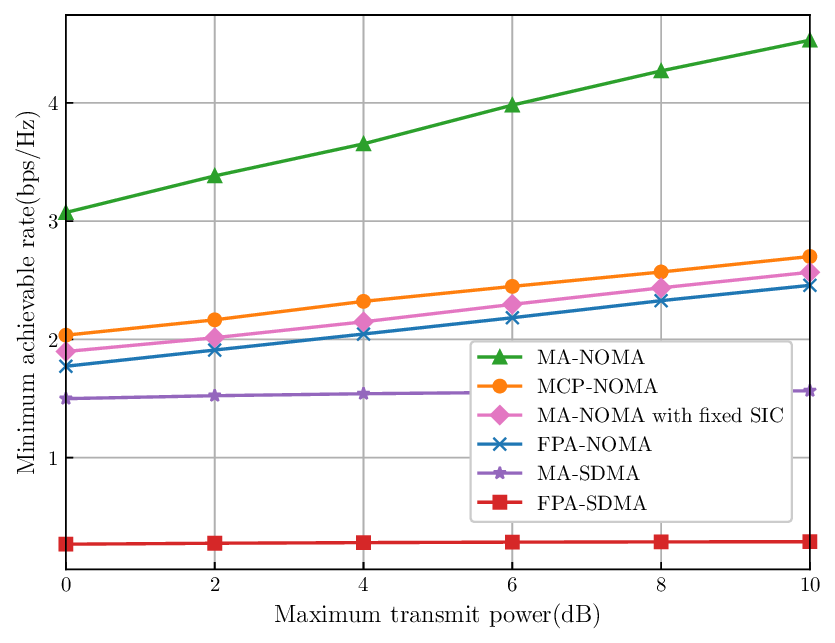}
        \caption{Minimum achievable rates for different schemes versus maximum transmit power.}
        \label{power vs rate}
    \end{minipage}
    \hfill
    \begin{minipage}{0.49\linewidth}
        \centering
        \includegraphics[width=3in]{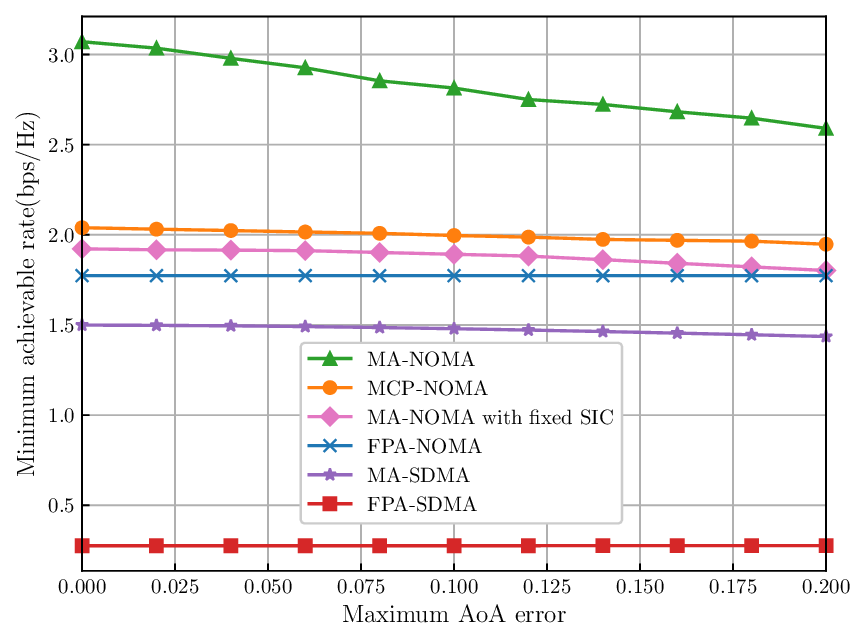}
        \caption{Impact of the AoA error on the performance of the proposed
algorithms.}
        \label{aoa vs rate}
    \end{minipage}
\end{figure*}

Fig. \ref{power vs rate} compares the minimum achievable rate of the different schemes versus the maximum transmit power of the BS. It can be seen that the minimum achievable rate of the NOMA-based schemes increases remarkably as the maximum transmit power increases.
However, the SDMA-based schemes show less improvement because increasing the transmit power can not only improve the received SNR at each user, but also increase the interference among multiple users. This challenge persists when the number of antennas is fewer than the number of users, limiting SDMA-based schemes' ability to effectively eliminate interference.
%In addition, the optimization problem aims to maximize the minimum achievable rate among multiple users, which is subject to user fairness constraint. Therefore, it cannot improve the performance by allocating more power to certain users, e.g., in the sum rate problem, we can allocate as much power as possible to the user with the best channel quality while ensuring that the other users satisfy the minimum rate requirement \cite{4zhu2019joint}. 
%Furthermore, our proposed MA-NOMA scheme outperforms all other schemes in the minimum achievable rate and its minimum achievable rate increases almost linearly, indicating that our proposed algorithm always finds the same positions for deploying the MAs.

\subsection{Impact of Imperfect FRI}
All the above simulation results are based on perfect FRI. However, perfect FRI estimation is challenging in real communication systems. Therefore, we need to verify the performance of the proposed algorithm under imperfect FRI. For this purpose, we introduce angle estimation error and path response coefficient estimation error to evaluate the proposed algorithm. Specifically, we define the AoAs error as the difference between the actual AoAs and the estimated AoAs. Let $\hat \vartheta _{k,i}^r$, $\hat \varphi _{k,i}^r$ and $\hat \omega _{k,i}^r$ represent the estimated AoAs of $i$-th channel path between the BS and the user $k$, where $1 \le k \le K$, $1 \le i \le L$.  The differences between the actual AoAs and the estimated AoAs are denoted as random variables following uniform distributions, i.e., $\vartheta _{k,i}^r - \hat \vartheta _{k,i}^r \sim {{\cal U}}\left[ {{{ - \mu } \mathord{\left/{\vphantom {{ - \mu } 2}} \right. \kern-\nulldelimiterspace} 2},{\mu  \mathord{\left/
{\vphantom {\mu  2}} \right.\kern-\nulldelimiterspace} 2}} \right]$, $\varphi _{k,i}^r - \hat \varphi _{k,i}^r \sim {{\cal U}}\left[ {{{ - \mu } \mathord{\left/{\vphantom {{ - \mu } 2}} \right. \kern-\nulldelimiterspace} 2},{\mu  \mathord{\left/{\vphantom {\mu  2}} \right.
\kern-\nulldelimiterspace} 2}} \right]$ and $\omega _{k,i}^r - \hat \omega _{k,i}^r \sim {{\cal U}}\left[ {{{ - \mu } \mathord{\left/
 {\vphantom {{ - \mu } 2}} \right.\kern-\nulldelimiterspace} 2},{\mu  \mathord{\left/{\vphantom {\mu  2}} \right.\kern-\nulldelimiterspace} 2}} \right]$, where $\mu$ is maximum AoA error. Moreover, the PRM error is denoted as the difference between the actual PRMs and estimated PRMs. The estimated channel path response coefficient is ${{{\hat \sigma }_i}}$, and the difference between $\sigma_i$ and ${\hat \sigma}_i$ is a random variable following the CSCG distribution, i.e., $\frac{{{\sigma _i} - {{\hat \sigma }_i}}}{{\left| {{\sigma _i}} \right|}} \sim {{\cal C}{\cal N}}\left( {0,\nu } \right)$, $1 \le i \le L$, where the $\nu$ is the variance of the normalized PRM error.

 Fig. \ref{aoa vs rate} illustrates the impact on the system performance in the presence of errors in the estimation of AoAs. In order to fairly verify the performance in the face of imperfect FRI, we first assume that the AoAs are perfectly estimated and obtain the optimized MAs' positions through Algorithm \ref{algorithm u}. Then, we impose an error on the estimated AoAs and recalculate the channel response matrix.
 % in accordance with the optimized MAs' positions obtained in the previous step. 
 % Finally, we calculate the max-min achievable rate through Algorithm \ref{algorithm AO}. 
 From the simulation results, the AoA error affects the MA-aided schemes as well as the MCP-NOMA scheme, in which the MA-NOMA is more suffered. However, its performance is still significantly better than the other schemes, which indicates that our algorithm is robust.

\begin{figure}[h!]
\centering
\includegraphics[width=3in]{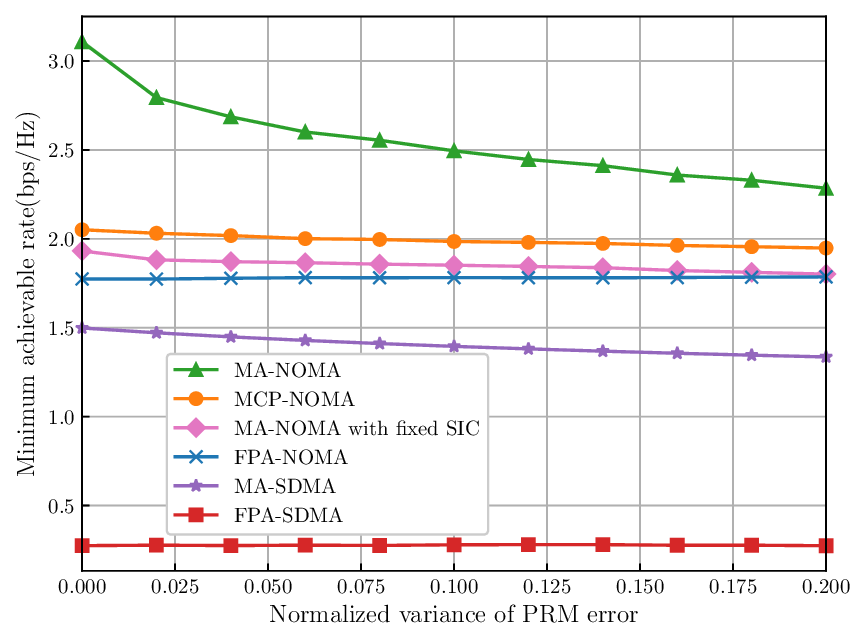}
\caption{Impact of the PRM error on the performance of the proposed
algorithms.}
\label{prm vs rate}
\end{figure}

Fig. \ref{prm vs rate} shows the impact of PRM error on the performance of the proposed algorithm. Similarly, we first assume that the PRM estimation is perfect, and based on that, we obtain the optimized MAs' positions. Then, we calculate the actual channel response matrix. 
% Furthermore, we optimize the precoding matrix and the decoding indicator matrix iteratively using Algorithm \ref{algorithm AO} in order to get the max-min achievable rate. 
Simulation results show that the PRM error impairs the performance of MA-based schemes and the MCP-NOMA scheme. 
% The minimum achievable rate decreases further as the PRM error increases.
However, our proposed scheme still outperforms all other schemes.
% Moreover, PRM error has a greater impact on the performance of the algorithm than AoA error. 
% % It is worth noting that the MA-NOMA with fixed SIC scheme still exhibits better performance than the FPA-NOMA scheme in the two imperfect FRI cases, which shows that even though the MA-NOMA with fixed SIC scheme does not have flexible decoding scheme, MA can help the communication system to achieve higher reliability in the presence of channel estimation errors compared to the conventional FPA system.

\section{Conclusion}

In this paper, we studied the MA-aided downlink NOMA transmission from a BS with FPAs to multiple users each with a single MA. 
% The MA-aided communication system can significantly increase the minimum achievable rate among multiple users compared to the conventional FPA-equipped system. 
We first modeled the downlink channel from the BS to multiple users as a function of the APV of users' MAs. 
Then, to overcome the shortcomings of the conventional NOMA with the fixed decoding scheme, we proposed an adaptive SIC decoding method to flexibly adjust the decoding indicator matrix of multiple users.
% In this flexible decoding scheme, a matrix consisting of binary optimization variables was used to indicate the decoding scheme for each user, which was called the decoding indicator matrix. In addition, we rewrote the calculation form of achievable rate based on the decoding indicator matrix.
Furthermore, we formulated an optimization problem to maximize the minimum achievable rate among multiple users by jointly optimizing the positions of MAs, the precoding matrix, and the decoding indicator matrix, subject to the constraints of limited MA movable region, maximum transmit power of the BS, and the SIC decoding condition.
In order to solve this non-convex problem with highly coupled variables, we proposed a two-loop iterative algorithm that combines improved HO method with AO method.
% In the inner loop, we used the AO algorithm to alternately optimize the precoding matrix and the decoding indicator matrix. In the outer loop, an improved HO algorithm was exploited to optimize the MAs' positions.
Simulation results showed that our proposed algorithm exhibited significant advantages over both FPA-based schemes and other benchmark schemes. 
% Moreover, the adaptive decoding scheme enabled our proposed algorithm to demonstrate better performance for any number of antennas and users. 
% Finally, we also investigated the performance of the algorithm under imperfect FRI estimation scenarios, and the results illustrated that although estimation errors had an impact on the performance, our proposed algorithm still outperformed all other benchmark schemes.

\bibliographystyle{IEEEtran}
\bibliography{mybibi}
\end{document}